\documentclass{aa}
\usepackage{color,graphicx}
\usepackage{amsmath} 
\usepackage[varg]{txfonts} 
\usepackage{multirow}
\usepackage{natbib}
\bibpunct[]{(}{)}{;}{a}{}{,} 

\begin{document}

\title{Solar meridional circulation from twenty-one years of\\
SOHO/MDI and SDO/HMI observations\thanks{
  The long-term averages of the measured travel-time shifts are available in electronic form at the CDS via anonymous ftp to \texttt{cdsarc.u-strasbg.fr (130.79.128.5)} or via \texttt{http://cdsweb.u-strasbg.fr/cgi-bin/qcat?J/A+A/}}
}
\subtitle{Helioseismic travel times and forward modeling in the ray approximation}
\titlerunning{Solar meridional circulation}

\author{
Zhi-Chao Liang \inst{\ref{mps}} \and
Laurent Gizon \inst{\ref{mps},\ref{gottingen},\ref{NYUAD}} \and
Aaron C. Birch \inst{\ref{mps}} \and
Thomas L. Duvall, Jr. \inst{\ref{mps}} \and
S.~P. Rajaguru \inst{\ref{iiap}}
}

\institute{
Max-Planck-Institut f\"ur Sonnensystemforschung, Justus-von-Liebig-Weg 3, 37077 G\"ottingen, Germany\\
\email{zhichao@mps.mpg.de} \label{mps}\and
Institut f\"ur Astrophysik, Georg-August-Universit\"at G\"ottingen, Friedrich-Hund-Platz 1, 37077 G\"ottingen, Germany \label{gottingen}\and
Center for Space Science, NYUAD Institute, New York University Abu Dhabi, PO Box 129188, Abu Dhabi, UAE \label{NYUAD}\and
Indian Institute of Astrophysics, Koramangala II Block, Bangalore, India \label{iiap}
}

\date{Received 19 June 2018 / Accepted 21 August 2018}

\abstract
{
  The solar meridional flow is an essential ingredient in flux-transport dynamo models.
  However, no consensus on its subsurface structure has been reached.
} {
  We merge the data sets from SOHO/MDI and SDO/HMI with the aim of achieving a greater precision on helioseismic measurements of the subsurface meridional flow.
} {
  The south-north travel-time differences are measured by applying time-distance helioseismology to the MDI and HMI medium-degree Dopplergrams covering May 1996--April 2017.
  Our data analysis corrects for several sources of systematic effects: $P$-angle error, surface magnetic field effects, and center-to-limb variations.
  For HMI data, we used the $P$-angle correction provided by the HMI team based on the Venus and Mercury transits.
  For MDI data, we used a $P$-angle correction  estimated  from the  correlation of MDI and HMI data during the period of overlap.
  The center-to-limb effect is estimated from the east-west travel-time differences and is different for MDI and HMI observations.
  An interpretation of the travel-time measurements is obtained using a forward-modeling approach in the ray approximation.
} {
  In the latitude range 20$\degr$--35$\degr$, the travel-time differences are similar in the southern hemisphere for cycles 23 and 24.
  However, they differ in the northern hemisphere between cycles 23 and 24.
  Except for cycle 24's northern hemisphere, the measurements favor a single-cell meridional circulation model where the poleward flows persist down to $\sim$0.8~$R_\odot$, accompanied by local inflows toward the activity belts in the near-surface layers.
  Cycle 24's northern hemisphere is anomalous: travel-time differences are significantly smaller when travel distances are greater than 20$\degr$.
  This asymmetry between northern and southern hemispheres during cycle 24 was not present in previous measurements, which assumed a different $P$-angle error correction where south-north travel-time differences are shifted to zero at the equator for all travel distances.
  In our measurements, the travel-time differences at the equator are zero for travel distances less than $\sim$30$\degr$, but they do not vanish for larger travel distances.
  This equatorial offset for large travel distances need not be interpreted as a deep cross-equator flow; it could be due to the presence of asymmetrical local flows at the surface near the end points of the acoustic ray paths.
} {
  The combined MDI and HMI helioseismic measurements presented here contain a wealth of information about the subsurface structure and the temporal evolution of the meridional circulation over 21 years.
  To infer the deep meridional flow, it will be necessary to model the contribution from the complex time-varying flows in the near-surface layers.
}


\maketitle

\section{Introduction}
We define solar meridional circulation as the axisymmetric component of the meridional flow in a spherical-polar coordinate system where the polar axis coincides with the Sun's rotation axis.
We assume in this definition that all layers rotate about a single rotation axis.
The surface meridional flow, first measured by \citet{Duvall1979}, is poleward with a peak speed of 10--20~m~s$^{-1}$ at low- and mid-latitudes, and is known to persist through the convection zone to some extent \citep{Giles1997}.
It has been found by various methods that the magnitude and profile of the meridional flow varies with solar activity level \citep[e.g.,][]{Chou2001,Haber2002,Beck2002,Zhao2004,Hathaway2010,Ulrich2010,Liang2015b,Komm2015}.
In parallel, a pattern of inflows toward the active regions develops \citep[e.g.,][]{Gizon2001,Loeptien2017} and moves equatorward in step with the activity belts as the cycle progresses.
It is believed that these inflows, which in general have a component in the meridional plane, account for a significant part of the cyclic change of the meridional flow \citep{Gizon2004b,Svanda2008}.
In addition, the meridional flow is capable of transporting magnetic flux and is considered a likely mechanism in flux-transport dynamo models for the conveyance of the field in the solar interior \citep[see, e.g., recent review by][]{Cameron2017}.

Recently, a number of inconsistent helioseismic results were reported \citep{Zhao2013,Schad2013,Jackiewicz2015,Rajaguru2015,Boening2017,Chen2017,Lin2018}.
The main reason is that helioseismic measurements suffer from a variety of systematic errors such as instrumental misalignment, the center-to-limb variation, and the influence of the surface magnetic field \citep[e.g.,][]{Beck2005,Duvall2009,Zhao2012,Liang2015a}.
Furthermore, the meridional flow is more than one order of magnitude weaker than other major flows inside the Sun.
Using the noise model by \citet{Gizon2004}, it has been suggested by \citet{Braun2008} and by \citet{Hanasoge2009} that an estimate of the meridional flow at the bottom of the convection zone at a level of precision of 1~m~s$^{-1}$ requires tens of years of data.
Thus, helioseismic measurements of a deep meridional flow are in need of a very long observational time series.

The Michelson Doppler Imager on board the Solar and Heliospheric Observatory \citep[SOHO/MDI:][]{Scherrer1995} and the Helioseismic and Magnetic Imager on board the Solar Dynamical Observatory \citep[SDO/HMI:][]{Scherrer2012,Schou2012} have accumulated nearly two solar cycles of full-disk Doppler observations since May 1996.
Lately \citet{Liang2017}, taking care of most of the major systematics, performed a detailed comparison of the travel-time measurements of meridional circulation from the two data sets in a concurrent period from May 2010 to April 2011 and obtained a remarkable degree of consistency.
Building upon the previous success, we merge the two data sets in this work with hopes of tying down the meridional flow profile below 0.9~$R_\odot$.
In Sect.~\ref{sec:data}, we describe the data preparation and analysis.
In Sect.~\ref{sec:obs}, we present the measured travel-time shifts.
In Sect.~\ref{sec:model}, we model not only the global-scale meridional circulation that covers all latitudes, but also the inflows around the mean active latitudes in the upper convection zone as a separate component.
The measured travel-time shifts are compared with the forward-modeled travel-time shifts in Sect.~\ref{sec:cmpr}
We summarize in Sect.~\ref{sec:sum} with a discussion of the implications of the work presented here.

\section{Data and analysis} \label{sec:data}

\begin{figure}
  \resizebox{\hsize}{!}{\includegraphics{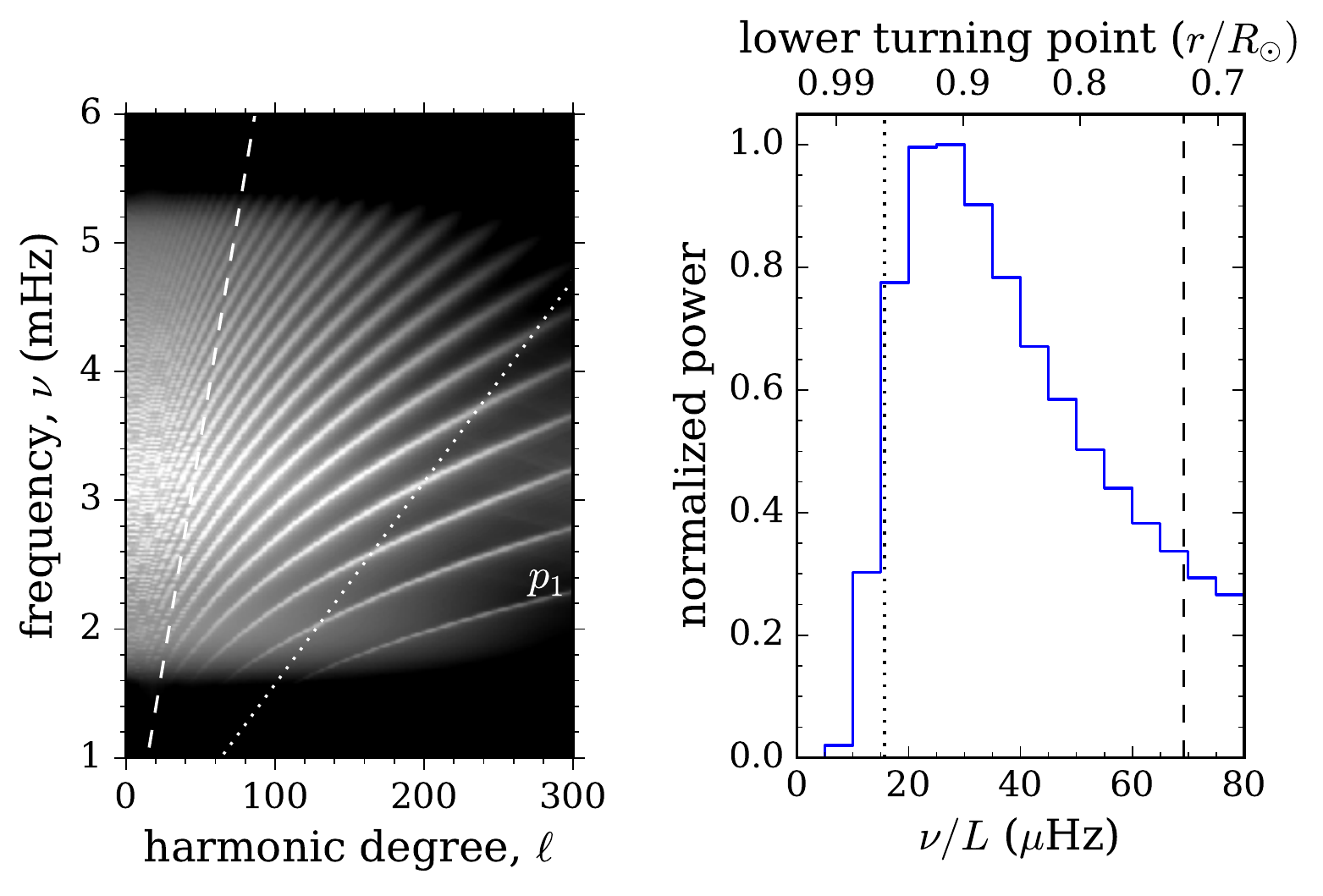}}
  \caption{ \label{fig:lnu}
    \emph{Left}: example $m$-averaged power spectrum from filtered and tracked HMI data as a function of frequency $\nu$ and harmonic degree $\ell$.
    To reduce random noise, an average is carried out over 30 individual daily power spectra observed in June 2010.
    The first visible ridge at low frequencies is the $p_1$ modes.
    \emph{Right}: histogram of mode power (blue) from the left panel as a function of $\nu/L$, where $L^2 = \ell(\ell+1)$.
    The corresponding radii of lower turning points from the ray approximation are indicated at the top.
    The dotted and dashed lines in both panels correspond to the modes that contribute most to our shallowest and deepest measurements in this work, respectively.
  }
\end{figure}

We use the medium-$\ell$ Dopplergrams spanning from 1 May 1996 to 30 April 2010 made by SOHO/MDI at a cadence of 60~s and from 1 May 2010 to 30 April 2017 made by SDO/HMI at a cadence of 45~s.
They are retrieved from the \textsf{mdi.vw\_v} and \textsf{hmi.vw\_v\_45s} data series in the Joint Science Operations Center\footnote{Joint Science Operations Center, \texttt{http://jsoc.stanford.edu/}} (JSOC) data system.
These medium-$\ell$ Dopplergrams are produced by smoothing and subsampling the full-resolution line-of-sight velocity images \citep{Kosovichev1997a}.
They have an image scale of $\sim$10~arcsec~pixel$^{-1}$ and are sensitive to oscillation modes in the regime where spherical harmonic degree $\ell \leq 300$.
Figure~\ref{fig:lnu} shows an example of an oscillation power spectrum computed from one-month HMI data.
Clearly, the power drops rapidly for $p$~modes that reside in a shallow subsurface layer (above 0.96~$R_\odot$).
That is, the medium-$\ell$ Dopplergrams contain little information about the near-surface layers, which is a direct outcome of the smoothing and resampling procedures.

\begin{figure*}
  \sidecaption
  \includegraphics[width=12cm]{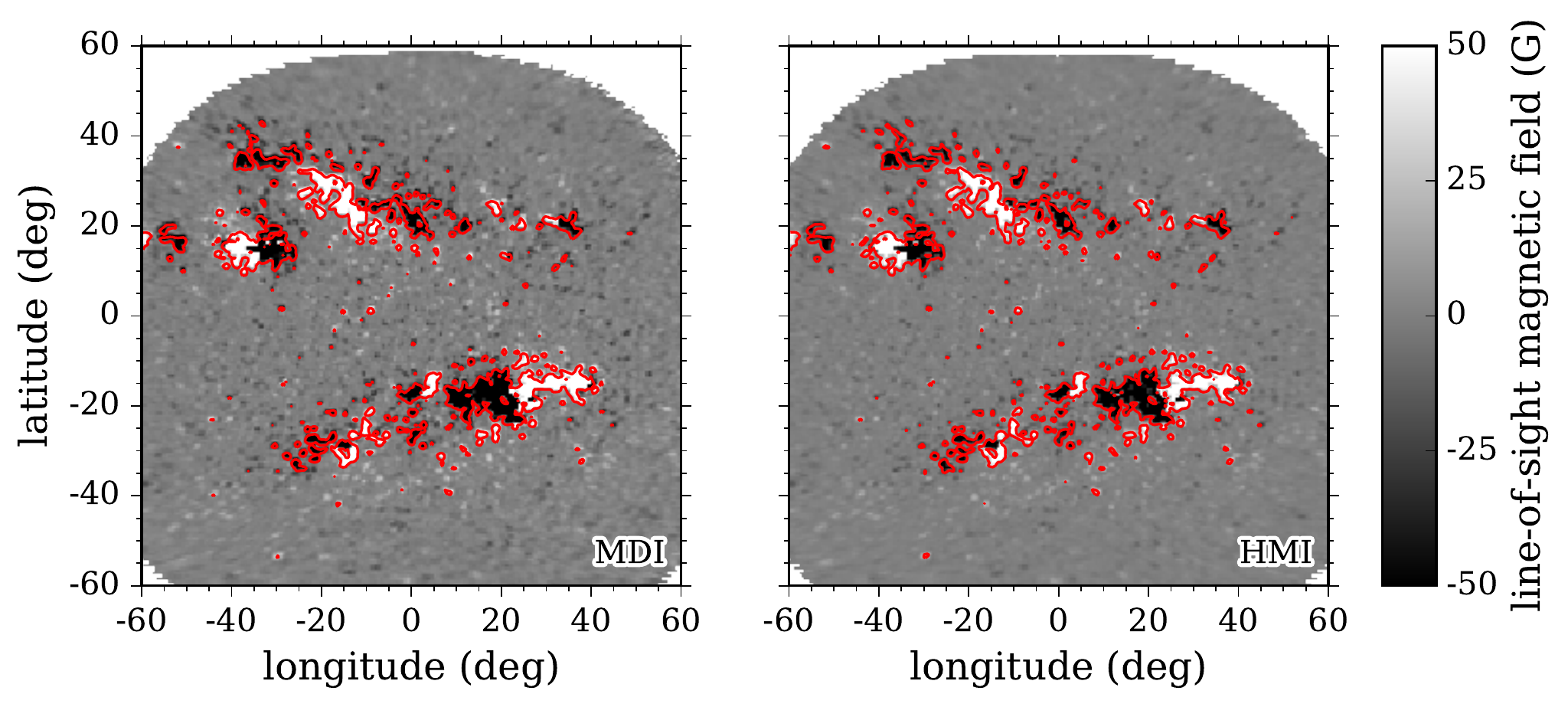}
  \caption{ \label{fig:bmask}
    Example magnetograms derived from MDI data (left) and HMI data (right) on 30 March 2011.
    The overlying red contours indicate the border of the masked areas determined by a threshold value of 30~G (left) and 20~G (right), inside which the data points are not used in the travel-time measurements.
    The MDI data on this day shown here is only for comparison since HMI data are used instead of MDI data after 1 May 2010 in this work.
  }
\end{figure*}

\subsection{Data reduction}\label{sec:reduction}

We select dates with a duty cycle of more than 75\%.
Because the SOHO spacecraft operations and MDI instrument events are not properly reflected in the MDI quality bits, the MDI event table\footnote{MDI Events Query Form,\\ \texttt{http://mditlm.stanford.edu/events/events.html}} is used to identify unusable images.
These events consist of operations such as emergency Sun reacquisitions, momentum managements, and pointing adjustments.
Also, any day in which the MDI instrument's focus, tuning, or exposure changes occurred is rejected.
There are corrupted images in MDI data series that are not indicated by the event table or the quality bits but can be picked out by inspecting the statistical keywords, \texttt{DATAMEAN} and \texttt{DATARMS}.
Straight lines are fitted to time series of the two keywords in each day by minimizing absolute deviation \citep[e.g.,][, Sect.~15.7]{Press1992} rather than the chi-square merit function to avoid the undesired sensitivity to the outliers.
MDI images whose statistical keyword values deviate from the daily fitted line by seven times the median of the absolute deviations are rejected.

\citet{Liang2017} reported that the travel-time shifts measured from MDI data give inconsistent results when the SOHO spacecraft was rotated by 180$\degr$.
Accordingly, we only select MDI images whose nominal position angle ($P$-angle) is zero (i.e., keyword \texttt{CROTA2} = 0).
As a result, half of the data from mid-2003 to mid-2010 are not used.
Because the SOHO spacecraft always flipped in the periods when the solar tilt angle $B_0$ is small, the selected MDI images since mid-2003 all have a large $B_0$ angle and thus the noise level of the travel-time measurement from MDI data is greater, at higher latitudes in particular.

A large-scale background signal owing to solar rotation and the orbital motion of the spacecraft is removed by subtracting a one-hour running mean for each pixel.
Furthermore, a band-pass frequency filter is applied to remove the solar convection signals at low frequencies (e.g., granulation and supergranulation) and the propagating-wave signals above the cutoff frequency.
The filter used has a flat top within 2--5~mHz and is tapered to zero by a half-cycle raised-cosine function of width 0.5~mHz at both ends.
These filtered images are mapped into heliographic coordinates using equidistant cylindrical projection with a map scale of 0.6~heliographic degree pixel$^{-1}$ and tracked with the Carrington rotation rate on a daily basis.
The interpolation method used in the mapping procedure is the bicubic spline \citep[e.g.,][, Sect.~3.6]{Press1992}.

It has been known that there is a $P$-angle error in MDI data due to instrumental misalignment \citep{Giles1997,Giles2000,Beck2005,Hathaway2010}.
More recently, \citet{Schuck2016} and \citet{Liang2017} compared contemporaneous images from MDI and HMI data sets, and found a time-varying $P$-angle offset between 0.18$\degr$ and 0.24$\degr$.
Since there is no information about the variation of this error before May 2010, to zeroth order, a constant value of 0.2$\degr$ is added to MDI keyword \texttt{CROTA2} along with the mapping procedure to correct the MDI $P$-angle error.
As for the $P$-angle of HMI, we note that a correction of $\sim$0.07$\degr$, based on the observations of the Venus and Mercury transits \citep{Couvidat2016,Hoeksema2018}, had been applied to the keyword \texttt{CROTA2} for the HMI images used here, and is taken into account in our mapping procedure.

To circumvent a systematic effect caused by the surface magnetic field on travel-time measurements \citep{Liang2015a}, we prepare the magnetograms for masking the active regions.
The line-of-sight magnetic field data are retrieved from the \textsf{mdi.fd\_m\_96m\_lev182} and \textsf{hmi.m\_45s} data series in the same period as for the Dopplergrams.
For simplicity, we process the two data sets at intervals of 96~min although the original HMI magnetograms have a high cadence of 45~s.
The MDI magnetograms are binned by a factor of 5 while the HMI magnetograms are binned by a factor of 20 to reduce the spatial resolution to that of the medium-$\ell$ data.
These binned magnetograms are tracked and remapped in the same way as for the Dopplergrams.
The tracked magnetograms are then averaged within a day at each pixel by taking the median instead of the mean to exclude cosmic-ray hits.
As the resulting magnetic field strength of each pixel is subject to factors like the image scale and cadence, binning procedure, and the tracking rate, we compute the standard deviation of pixels within 30~heliocentric degrees from the disk center on a quiet day available for the two data sets (i.e., 17 May 2010).
The threshold used for identifying the active regions is then determined as five times the standard deviation, namely, 30~G for MDI data and 20~G for HMI data.
An example of magnetograms derived from both data sets on the same day with masked areas overlaid is shown in Fig.~\ref{fig:bmask}.
The masked areas account for up to 25\% of data pixels at active latitudes in the solar maximum of cycle 23 and 22\% in the maximum of cycle 24 \citep[cf.][, Fig.~2]{Liang2015a}.

\subsection{Time-distance analysis}
To measure the travel-time shift arising from a subsurface flow in the meridional plane, we adopt a time-distance measurement scheme from \citet{Duvall2003}.
The cross-covariance function (CCF) is computed between time series of Doppler signals from pairs of points, separated by an angular distance $\Delta$, on opposing arcs aligned in the north-south direction.
Wave paths connecting pairs of points on the opposing arcs at the surface all intersect at a central point beneath the surface.
The two arcs, both at an angular distance of $\Delta/2$ from the central point, subtend an angle of 30$\degr$.
The number of points on an arc is determined by rounding the ratio of the arc length to the image scale (i.e., 0.6$\degr$ in this work) to the nearest integer.
A schematic plot of the arc-to-arc geometry is provided in the top panel of Fig.~\ref{fig:arc}.
If any of the paired points is within the masked areas mentioned in Sect.~\ref{sec:reduction}, the CCF of this pair is discarded and not included in the later averaging.
The CCFs computed from pairs of points on opposing arcs are averaged and assigned to the central point.
The above procedure is repeated for different central points situated within a 30$\degr$-wide strip along the central meridian at intervals of 0.6$\degr$ in longitude and latitude, and for different travel distances $\Delta$ ranging from 6$\degr$ to 42$\degr$ in increments of 0.6$\degr$.
These arc-averaged CCFs are further averaged over longitude and over available days in each month.
The northward and southward travel times, $\tau_\text{n}$ and $\tau_\text{s}$, are measured by fitting a Gabor wavelet function to the monthly-averaged CCFs around the single-skip wavelets in the positive and negative time lags, respectively \citep{Kosovichev1997b,Duvall1997}.
The window function used to isolate the wavelet and the peak to which the phase time is fitted in the CCF are detailed in \citet[][, Appendix~A]{Liang2017}.
The south-north travel-time difference is then defined as $\delta\tau_\text{sn}=\tau_\text{s}-\tau_\text{n}$, which is sensitive to the subsurface meridional flow along the wave path connecting the corresponding pair of points.

Regarding the center-to-limb effect \citep{Zhao2012}, we calculate the east-west travel-time difference, $\delta\tau_\text{ew}$, by aligning the opposing arcs in the east-west direction, placing the central points within a 30$\degr$-wide strip along the equator, and averaging over latitude instead of longitude, but otherwise in the same way as for $\delta\tau_\text{sn}$.
The antisymmetric part of the $\delta\tau_\text{ew}$ about the equator, $\widetilde{\delta\tau}_\text{ew}$, is expected to represent the center-to-limb variation.
To take into account the annual variation of the $B_0$ angle to some extent \citep{GonzalezHernandez2006,Zaatri2006}, the center-to-limb correction to the $\delta\tau_\text{sn}$ of each month is implemented as
\begin{equation} \label{eq:sn-ew}
\delta\tau_\text{sn}'(\lambda, \Delta) = \delta\tau_\text{sn}(\lambda, \Delta) - \widetilde{\delta\tau}_\text{ew}(\phi-\overline{B}_0, \Delta),
\end{equation}
where $\phi$ and $\lambda$ are longitude and latitude, $\overline{B}_0$ is an average of the $B_0$ of the used data in the month of concern, and the interpolation method for shifting the $\widetilde{\delta\tau}_\text{ew}$ by $\overline{B}_0$ is cubic spline.
However, this empirical correction cannot fully account for the center-to-limb effect in that meridians are great circles while parallels are not (except for the equator), and thus the $\delta\tau_\text{sn}$ and $\delta\tau_\text{ew}$ may ``see'' different center-to-limb variation at higher latitudes.
If the $\delta\tau_\text{ew}$ were instead measured along great circles in the east-west direction, one might end up with a leakage of the meridional-flow signal into the $\delta\tau_\text{ew}$.

Finally, the monthly-corrected $\delta\tau_\text{sn}'$s are averaged over three periods, cycle 23 (May 1996--April 2008; 3051~days used), cycle 24 (May 2008--April 2017; 2833~days used), and both the cycles 23 and 24 (May 1996--April 2017), weighted by the number of days used in each month.
A total of 5884 days of data are analyzed.
To be on the safe side, we limit ourselves to the use of high-latitude areas such that the highest latitudes adopted in the long-term averages are the same for both hemispheres in each month even when the Sun was tilted.
The three measured $\delta\tau_\text{sn}'$ for cycle 23, cycle 24, and both cycles (denoted by $\delta\tau_\text{sn;23}'$, $\delta\tau_\text{sn;24}'$, and $\delta\tau_\text{sn;23+24}'$ for later convenience), along with the standard error of the temporal mean, are available at the CDS.

\section{Measured travel-time shifts} \label{sec:obs}

\begin{figure}
  \resizebox{\hsize}{!}{\includegraphics{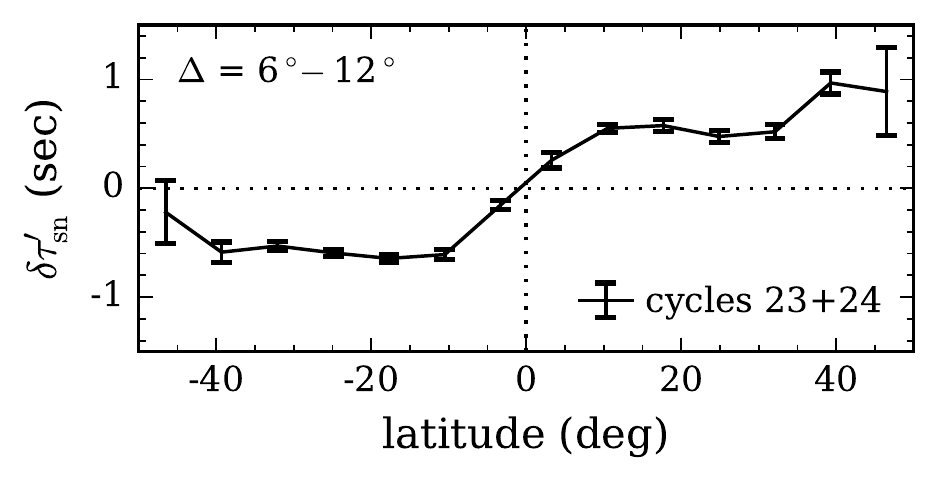}}
  \caption{ \label{fig:typ}
    Latitudinal profile of the 21-yr averaged measurement for short distances.
    The $\delta\tau_\text{sn;23+24}'$, averaged over the distance range $\Delta = 6\degr$--$12\degr$, is plotted as a function of latitude.
    The data values are binned every 7.2$\degr$ in latitude.
    The error bars give the standard deviation of the mean in each binning interval.
  }
\end{figure}

Figure~\ref{fig:typ} shows the latitudinal profile of the 21-yr averaged travel-time shifts, $\delta\tau_\text{sn;23+24}'$, for short distances.
The sense of the $\delta\tau_\text{sn}'$ is ``south minus north'' travel-time difference, so that a positive value of short-distance measurements indicates a northward flow in the near-surface layers.
The error bars at latitudes higher than 40$\degr$ are notably large because of the foreshortening effect and our conservative approach to the use of high-latitude areas.
The lack of small-$B_0$ data during the period from mid-2003 to mid-2010 exacerbates the problem.
In the following, we first examine the temporal variation of the measured travel-time shifts and then present more long-term averaged results.

\begin{figure}
  \resizebox{\hsize}{!}{\includegraphics{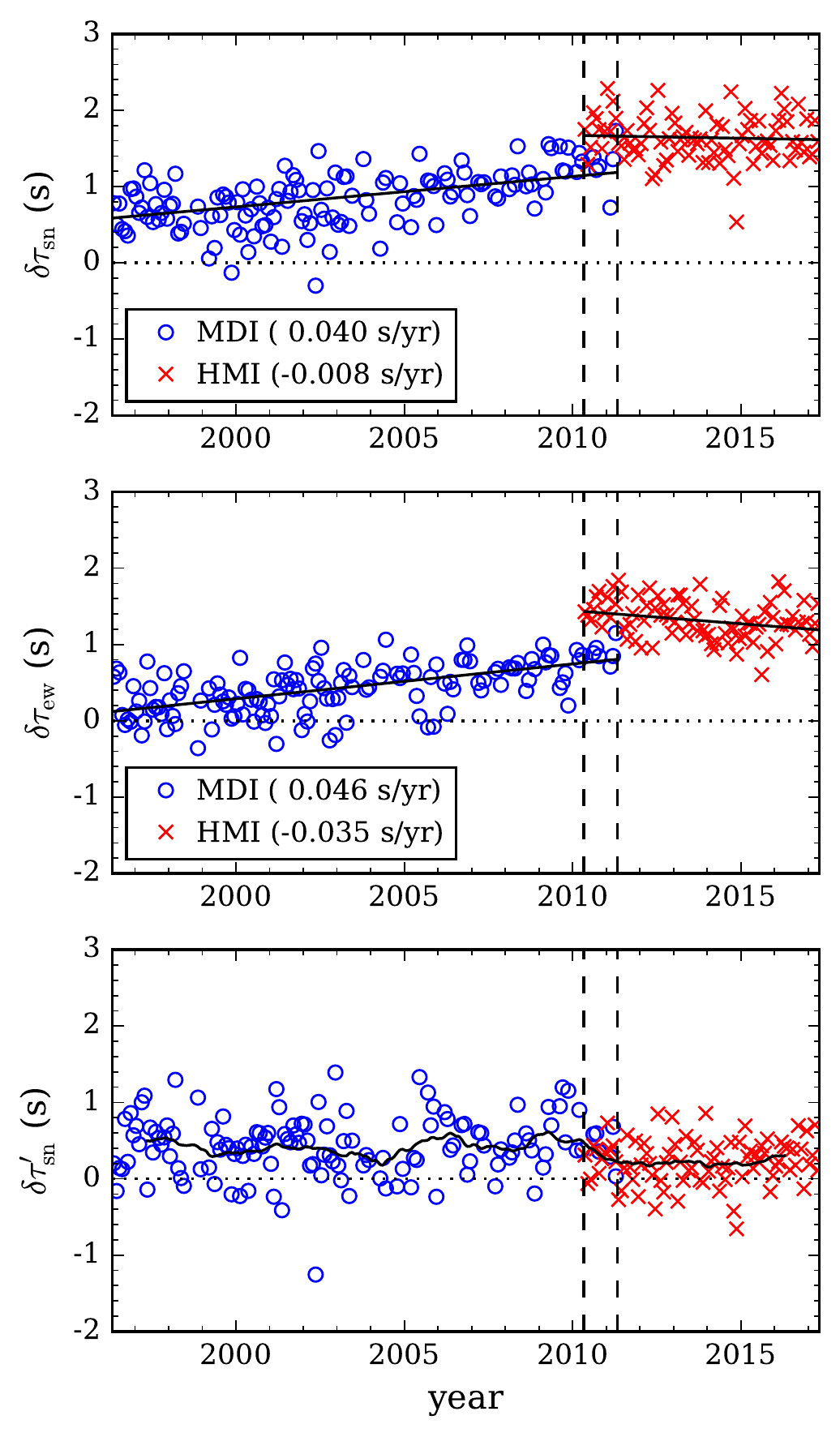}}
  \caption{ \label{fig:ctol}
    Temporal trend of the $\delta\tau_\text{sn}$ (\emph{top}), $\delta\tau_\text{ew}$ (\emph{middle}), and $\delta\tau_\text{sn}'$ (\emph{bottom}) from MDI (blue circles) and HMI (red crosses) data.
    These monthly-measured travel-time shifts are antisymmetrized about the equator (for $\delta\tau_\text{sn}$ and $\delta\tau_\text{sn}'$) or the central meridian (for $\delta\tau_\text{ew}$), and averaged over the latitude range 20$\degr$--35$\degr$ and the distance range $\Delta = 6\degr$--$42\degr$.
    The typical standard errors of the mean over the distance range for data points in the top, middle, and bottom panels are 0.26~s, 0.24~s, and 0.34~s, respectively.
    In between the two vertical dashed lines is the overlap between the MDI and HMI observations.
    The black solid lines in the top two panels are fitted straight lines for the two individual data sets, while the black solid curve in the bottom panel is a two-year running mean.
    The fitted slopes are indicated in the lower left corner of the top two panels.
    We note that the $\delta\tau_\text{ew}$ shown in the middle panel have not been shifted by $\overline{B}_0$ whereas the $\delta\tau_\text{sn}'$ in the bottom panel result from subtraction of the $\overline{B}_0$-shifted $\delta\tau_\text{ew}$ as described in Eq.~(\ref{eq:sn-ew}).
  }
\end{figure}

\begin{figure}[h!]
  \resizebox{\hsize}{!}{\includegraphics{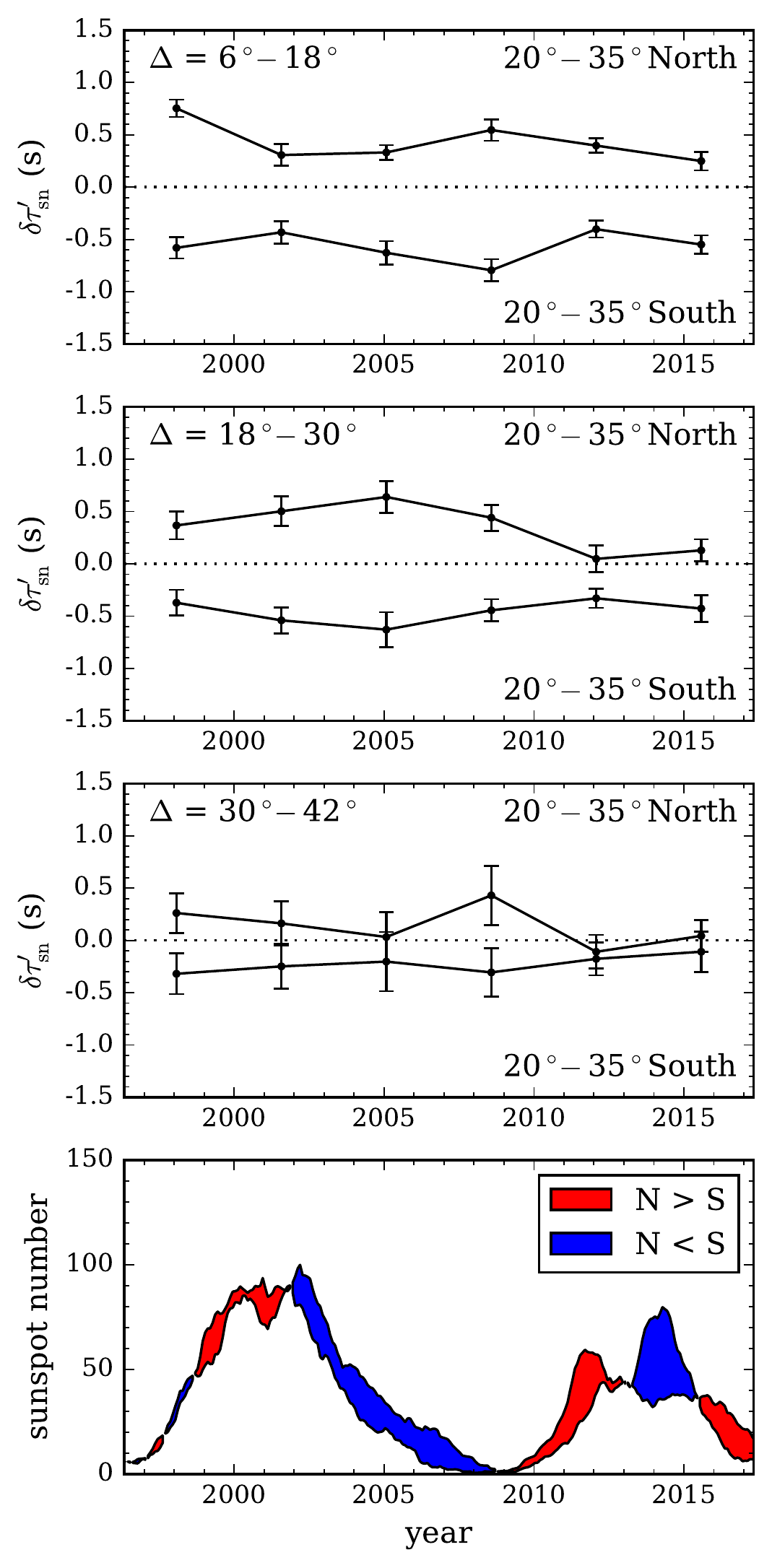}}
  \caption{ \label{fig:dt-vs-t}
    \emph{Top three}: temporal variations of the $\delta\tau_\text{sn}'$ in the northern and southern hemispheres for three distance ranges.
    The $\delta\tau_\text{sn}'$s are averaged over the latitude range 20$\degr$--35$\degr$ in the northern and southern hemispheres separately, and averaged over three distance ranges indicated in the top left corner of each panel.
    They are further binned every 3.5~years and the error bars give the standard deviation of the mean in each binning interval.
    \emph{Bottom}: 13-month running mean of the monthly sunspot numbers in the northern and southern hemispheres separately as a function of time.
    The shaded areas between the two curves indicate an excess of sunspot numbers in the northern hemisphere (red) or the southern hemisphere (blue).
  }
\end{figure}

\subsection{Long-term systematic error and solar cycle variation}
It has been reported by \citet{Liang2015b} that the center-to-limb effect measured from MDI data has a long-term variation.
They found that the $\delta\tau_\text{sn}$ and $\delta\tau_\text{ew}$ both have a gradually increasing trend; the larger the travel distance, the stronger the trend.
To remove this systematic trend, they fitted a straight line to the temporal trend of $\delta\tau_\text{ew}$ and then used the fitted slope to correct that of $\delta\tau_\text{sn}$.
Considering the accuracy required in this study, it would be important to scrutinize whether or not our monthly correction has accounted for it.

Figure~\ref{fig:ctol} shows the temporal trend of the monthly-measured $\delta\tau_\text{sn}$, $\delta\tau_\text{ew}$, and $\delta\tau_\text{sn}'$.
Evidently, the $\delta\tau_\text{sn}$ and $\delta\tau_\text{ew}$ from MDI data, averaged over the distance range $\Delta = 6\degr$--$42\degr$, both have a similar trend of $\sim$0.04~s per year at mid-latitudes.
This temporal trend may result in an enhancement of 0.44~s after 11 yr, which is an enormous effect for this type of measurement.
On the other hand, the $\delta\tau_\text{ew}$ from HMI data shows a decreasing trend as opposed to that from MDI data.
Besides, the difference between the trends of $\delta\tau_\text{sn}$ and $\delta\tau_\text{ew}$ from HMI data is greater than that from MDI data.
The reason might be that HMI started its observations from the rising phase of cycle 24 and the accumulated data have not yet covered a complete solar cycle.
As a result, the trend of $\delta\tau_\text{sn}$ from HMI data embodies the solar cycle variation from solar minimum to maximum, which is not present in the trend of $\delta\tau_\text{ew}$.

So far the physical origin of the center-to-limb effect is not fully understood.
Observations and numerical simulations suggested that the center-to-limb effect is related to a complex interaction of the solar dynamics and radiative transfer \citep{Zhao2012,Baldner2012,Kitiashvili2015,Schou2015,Chen2018}.
The long-term variation of the center-to-limb effect reported here implies that instrumental effects might also take part in it.

After removal of the center-to-limb variation, the trends of $\delta\tau_\text{sn}'$ from both data sets are overall consistent as shown in the bottom panel of Fig.~\ref{fig:ctol}.
However, the $\delta\tau_\text{sn}'$ from HMI data seems slightly smaller than that from MDI data.
To investigate this further, the temporal variations of the $\delta\tau_\text{sn}'$ at mid-latitudes of the northern and southern hemispheres are plotted separately for three distance ranges in the top three panels of Fig.~\ref{fig:dt-vs-t}, along with the sunspot number as a function of time in the bottom panel.

Two features in Fig.~\ref{fig:dt-vs-t} are noteworthy.
One is, unsurprisingly, the solar cycle variation of the $\delta\tau_\text{sn}'$.
Except the third panel for larger uncertainties, the $\delta\tau_\text{sn}'$s in the top two panels clearly show long-term variations which are, however, not in phase with each other.
In the first panel, the smaller amplitude of $\delta\tau_\text{sn}'$ (i.e., slower meridional flows) at mid-latitudes during the solar maxima is a known phenomenon and is attributed to the inflows toward the activity belts in the upper convection zone \citep[e.g.,][]{Zhao2004,Hathaway2011a}.
In the second panel, conversely, the amplitude of $\delta\tau_\text{sn}'$ reaches its maximum during the declining phase instead of the minimum phase as in the first panel.
As we will see later in Sect.~\ref{sec:model}, the local inflow structure in the upper convection zone produces an intricate pattern of travel-time shifts which alternates in latitude and in travel distance.
The different solar cycle variations of $\delta\tau_\text{sn}'$ that are not in phase for different distant ranges can be partly explained by the presence of the near-surface inflows.

The other noteworthy feature in Fig.~\ref{fig:dt-vs-t} is that the second and third panels show a much smaller amplitude of $\delta\tau_\text{sn}'$ in the northern hemisphere than that in the southern hemisphere during the rising and maximum phases of cycle 24.
Obviously, the slight reduction in $\delta\tau_\text{sn}'$ after 2010 in the bottom panel of Fig.~\ref{fig:ctol} comes from the larger distance measurements in the northern hemisphere.
This north-south asymmetry in the $\delta\tau_\text{sn}'$ during the rising and maximum phases of cycle 24 is so strong that it remains in the long-term average of the $\delta\tau_\text{sn}'$ as we shall see below.
It seems that the north-south asymmetry in the solar activity level is not the main cause of the north-south asymmetry in the $\delta\tau_\text{sn}'$ for larger distances since the $\delta\tau_\text{sn}'$ for shorter distances, which is expected to be affected more by the near-surface phenomena, does not show such a strong asymmetry.

\subsection{Long-term averaged results and north-south asymmetry} \label{sec:23vs24}

\begin{figure}[!h]
  \resizebox{0.97\hsize}{!}{\includegraphics{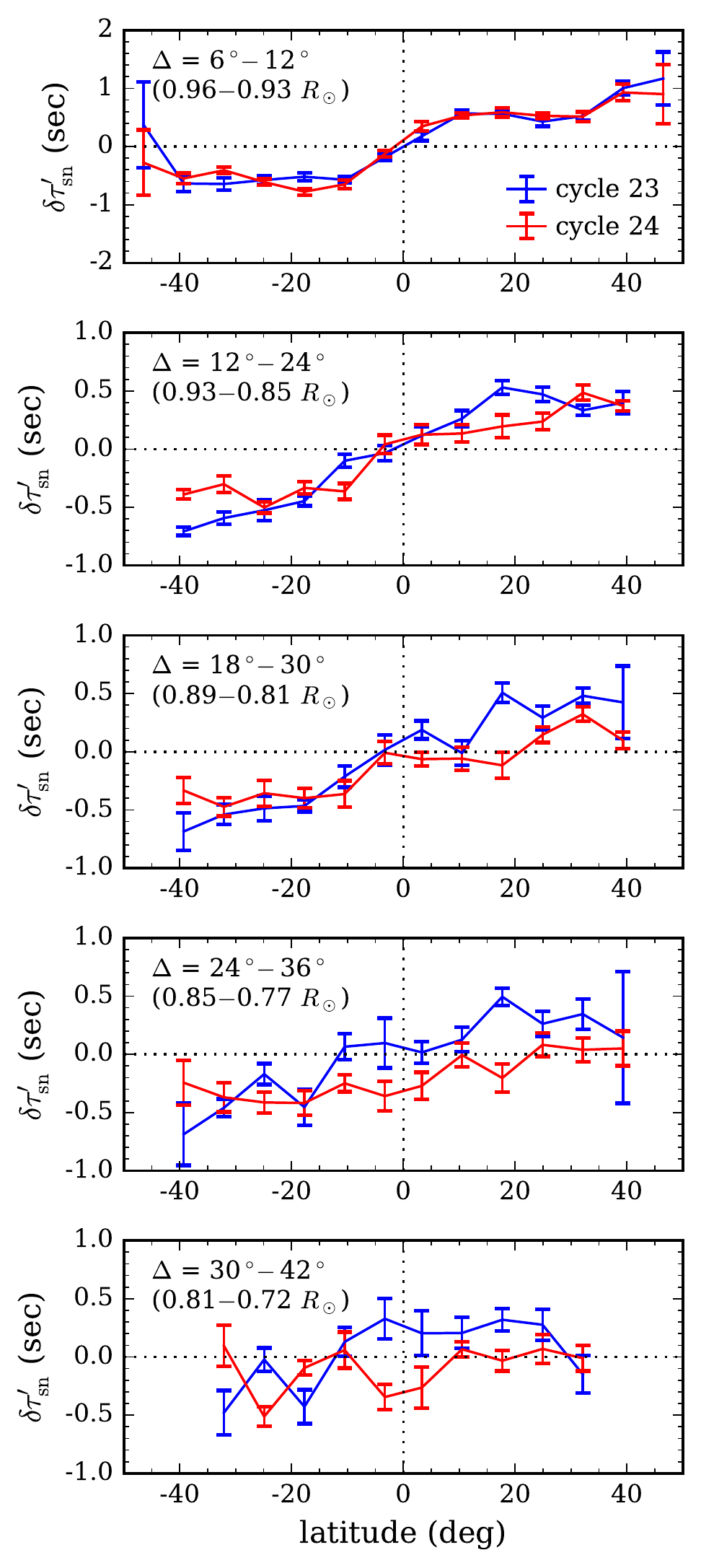}}
  \caption{ \label{fig:obs}
    Measured $\delta\tau_\text{sn}'$ for cycle 23 (blue) and cycle 24 (red), averaged over five distance ranges, as a function of latitude.
    The time periods are May 1996 to April 2008 for cycle 23 and May 2008 to April 2017 for cycle 24.
    The distance ranges, as well as the corresponding radii of the lower turning points from ray approximation, are indicated in the upper left corner of each panel; shortest at the top and largest at the bottom.
    The data values are binned every 7.2$\degr$ in latitude.
    The error bars give the standard deviation of the mean in each binning interval.
    We note that if we average the data over periods of seven years that cover the first halves of cycle 23 (May 1996 to April 2003; 2191~days used) and cycle 24 (May 2008 to April 2015; 2142~days used), we find essentially the same curves as shown here, however with a higher noise level.
  }
\end{figure}

The long-term averaged $\delta\tau_\text{sn;23}'$ and $\delta\tau_\text{sn;24}'$ are shown in Fig.~\ref{fig:obs}.
In the top panel, the shallowest measurements from both cycles are compatible with each other.
In contrast, the deeper measurements in the lower panels manifest substantial differences between the two cycles, particularly in the northern hemisphere.
Although the north-south asymmetry are present in both the $\delta\tau_\text{sn;23}'$ and $\delta\tau_\text{sn;24}'$, the degree of asymmetry in cycle 24 is so large that the $\delta\tau_\text{sn;24}'$ in the northern hemisphere vanishes in the bottom two panels.
As mentioned before, this reduction in the northern $\delta\tau_\text{sn;24}'$ mostly comes from the rising and maximum phases of cycle 24.

Also worth noting is the non-zero values at the equator for large-distance measurements.
As demonstrated by \citet[][, Fig.~4]{Liang2017}, a $P$-angle error in the data may introduce a systematic offset in the $\delta\tau_\text{sn}$ that gradually diminishes with increasing travel distance.
The reason why this systematic offset in $\delta\tau_\text{sn}$ is less noticeable for large distances is that the sound speed increases with depth and thus the $\delta\tau_\text{sn}$ becomes less sensitive to the leaking flows from solar rotation in a deeper layer.
In other words, a way to ascertain whether there is a $P$-angle error is to examine the measured value at the equator for short-distance cases, assuming there is no cross-equator flow in the near-surface layers.
The nearly-zero value of $\delta\tau_\text{sn;23}'$ at the equator as shown in the top panel of Fig.~\ref{fig:obs} indicates that for MDI data the $P$-angle correction is implemented to good effect, and the value of $\delta\tau_\text{sn;24}'$ at the equator, though slightly positive, is still within the error bar.
However, the values at the equator significantly deviate from zero for large-distance measurements, especially that of $\delta\tau_\text{sn;24}'$ as shown in the bottom two panels.

\begin{figure}
  \resizebox{\hsize}{!}{\includegraphics{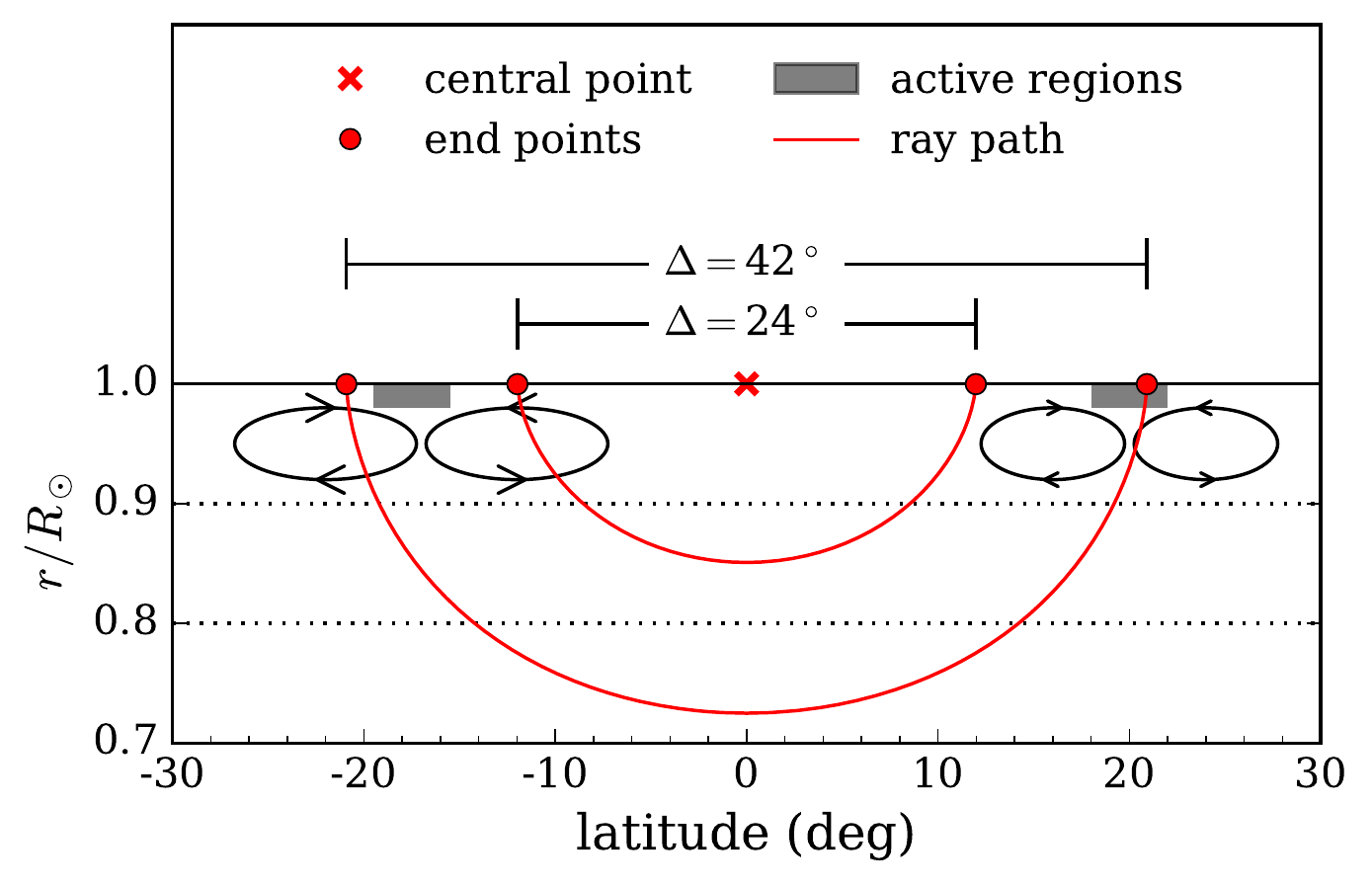}}
  \caption{ \label{fig:ray}
    Sketch for a possible scenario where the $\delta\tau_\text{sn}$ for large distances have non-zero values at the equator in the absence of a deep cross-equator flow.
    While the target location (cross symbol) is at the equator, the points (filled circles) that are connected by the acoustic ray paths (red curves) are located around the activity belts for large travel distances.
    The ellipses with arrow heads depict local flows around active regions.
    If these local flows are not the same in both hemispheres, they will contribute to the $\delta\tau_\text{sn}$ at these large travel distances.
    However, no contribution is expected at short travel distances (say $< 20\degr$) for which the end points do not reach the activity belts.
  }
\end{figure}

\label{sec:dP}
Since the $P$-angle-induced travel-time shift has its largest effect for short-distance measurements, the non-zero values for large-distance measurements at the equator are unlikely to be caused by an instrumental misalignment.
While a possible cross-equator flow is not excluded, these non-zero values at the equator for large-distance measurements might be caused by the north-south asymmetry of the near-surface inflows because the pairs of points for large-distance measurements at the equator are located around the active latitudes, as sketched in Fig.~\ref{fig:ray}.
Therefore, a non-zero $\delta\tau_\text{sn}'$ at the equator for large distances need not be interpreted as a cross-equator flow in the deep convection zone.
A correct interpretation would require an inversion or forward modeling that allows the presence of an asymmetric flow field, which is beyond the scope of this article.

We note that the results in \citet{Rajaguru2015} showed no such feature at the equator for large-distance measurements from HMI data, for they attributed it to a $P$-angle error and forced the measured values around the equator to be zero for all distances by shifting the $\delta\tau_\text{sn}$ with a distance-dependent value at all latitudes.
This might also be the reason why the strong north-south asymmetry for cycle 24 shown here was absent from their results.

\section{Forward modeling in the ray approximation} \label{sec:model}

Here we construct a few simple cellular flow models and compute their forward-modeled travel-time shifts as an aid to have an initial interpretation of the measurements.
It is important to stress that we do not attempt, in this very first step, to apply a fitting procedure for a best flow model, which certainly requires a thorough exploration of the model's parameter space.
Even so, we can still have a simple picture of solar meridional circulation through a forward-modeling study.

\subsection{Flow models}

Since this work is solely concerned with the measurement of flows in the meridional plane, we choose to study the flow field $\vec{u} = u_r(r,\theta)\,\vec{\hat{r}} + u_\theta(r,\theta)\,\vec{\hat{\theta}}$, where $r$ is the distance from the Sun's center, $\theta$ is the colatitude, and $\vec{\hat{r}}$ and $\vec{\hat{\theta}}$ are the unit vectors in the direction of increasing $r$ and $\theta$, respectively.
We focus our study on separable flow fields of the form
\begin{align}
    u_r(r,\theta) &= A_r f(r)g(\theta), \\
    u_\theta(r,\theta) &= A_\theta F(r)G(\theta), \label{eq:FG}
\end{align}
where $A_r$ and $A_\theta$ are scalar quantities.
The flow field conserves mass, that is $\nabla\cdot(\rho\vec{u})=0$, which enables us to relate $u_r(r,\theta)$ and $u_\theta(r,\theta)$ and obtain
\begin{align}
    A_r &= -A_\theta, \\
    F(r) &= \frac{1}{r\rho(r)}\frac{d}{dr} \Big( r^2\rho(r) f(r) \Big), \label{eq:df} \\
    g(\theta) &= \frac{1}{\sin\theta} \frac{d}{d\theta} \Big( \sin\theta \, G(\theta) \Big),
\end{align}
where $\rho(r)$ is the density profile taken from the solar model S \citep{Christensen-Dalsgaard1996}.

\begin{table}
  \caption{Parameters of flow models} 
  \centering
\begin{tabular}{c c c c c c c c}
  \hline\hline
  Name\rule{0pt}{2ex} & $u_\theta^\text{max}$ & $\theta_\text{n}$ & $\theta_\text{s}$ & $\alpha$ & $r_\text{t}$ & $r_\text{b}$ & $r_\text{c}$ \\
  & (m~s$^{-1}$) & & & & ($R_\odot$) & ($R_\odot$) & ($R_\odot$) \\
  \hline
  MC1 & 14.2 & 0$\degr$ & 180$\degr$ & 1 & 1 & 0.70 & 0.80 \\
  MC2 & 14.2 & 0$\degr$ & 180$\degr$ & 1 & 1 & 0.70 & 0.77 \\
  \multirow{2}{*}{MC3} & \multirow{2}{*}{14.2} & \multirow{2}{*}{0$\degr$} & \multirow{2}{*}{180$\degr$} & \multirow{2}{*}{1} & \multirow{2}{*}{1} & \multirow{2}{*}{0.70} & 0.92 \\
                       & & & & & & & 0.80 \\
  \hline
  \multirow{2}{*}{LC1} & \multirow{2}{*}{7.1} & 33$\degr$ & 113$\degr$ & \multirow{2}{*}{2} & \multirow{2}{*}{1} & \multirow{2}{*}{0.81} & \multirow{2}{*}{0.92} \\
                       &                          & 67$\degr$ & 147$\degr$ \\
  \hline
  \multirow{2}{*}{LC2} & \multirow{2}{*}{12.0} & 33$\degr$ & 113$\degr$ & \multirow{2}{*}{2} & \multirow{2}{*}{1} & \multirow{2}{*}{0.94} & \multirow{2}{*}{0.97} \\
                       &                          & 67$\degr$ & 147$\degr$ \\
  \hline
  \multirow{2}{*}{LC3} & \multirow{2}{*}{6.5} & 33$\degr$ & 113$\degr$ & \multirow{2}{*}{2} & \multirow{2}{*}{1} & \multirow{2}{*}{0.72} & \multirow{2}{*}{0.85} \\
                       &                          & 67$\degr$ & 147$\degr$ \\
  \hline
\end{tabular}
\tablefoot{ \label{tab:parm} 
  The MC models consist of a global cellular flow covering all latitudes, while the LC models consist of two local cellular flows located around the activity belts in the northern and southern hemispheres symmetrically.
  The $u_\theta^\text{max}$ refers to the maximal $u_\theta$ at the surface, and the $\alpha$ determines where the $u_\theta$ peaks in latitude (see Eq.~\ref{eq:g}).
  The parameters $r_\text{t}$, $r_\text{b}$, $\theta_\text{n}$, $\theta_\text{s}$ refer to the top, bottom, northern and southern boundaries of the cellular flow, respectively.
  The characteristic depth $r_\text{c}$ is the layer where the poleward flow transitions to the equatorward flow or vice versa.
}
\end{table}

\begin{figure*}[h!]
  \centering
  \includegraphics[width=17cm]{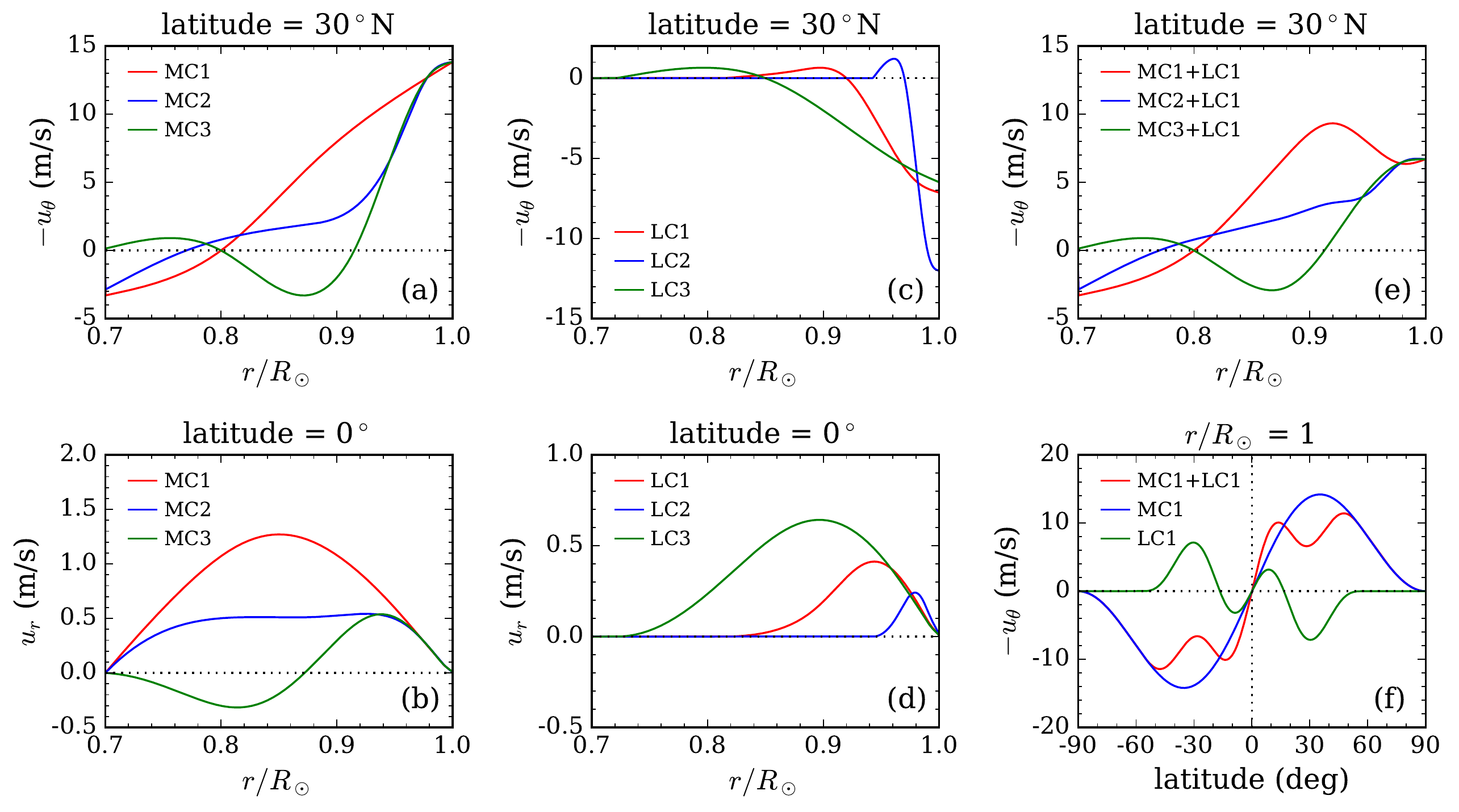}
  \caption{ \label{fig:ur+uh}
    Latitudinal and radial profiles of flow models listed in Table~\ref{tab:parm}.
    Panels (a) and (b) show vertical cuts through the $u_\theta$ and $u_r$ of the MC models as a function of $r$, respectively.
    Panels (c) and (d) show vertical cuts through the $u_\theta$ and $u_r$ of the LC models as a function of $r$, respectively.
    Panel (e) shows vertical cuts through the $u_\theta$ of the sum of MC models and model LC1 as a function of $r$.
    Panel (f) shows horizontal cuts at the surface through the $u_\theta$ of the sum of MC models and model LC1 as a function of latitude.
    We note that all the flow models in this work are antisymmetric about the equator.
  }
\end{figure*}

The latitudinal profile of the cellular flow model is given by
\begin{equation} \label{eq:g}
  G(\theta) = \left\{ \begin{array}{ll}
      \sin\left( 2\pi\frac{\theta - \theta_\text{n}}{\theta_\text{s} - \theta_\text{n}} \right) \sin^\alpha\left( \pi\frac{\theta-\theta_\text{n}}{\theta_\text{s}-\theta_\text{n}}\right) & \text{for}\; \theta_\text{n} \leq \theta \leq \theta_\text{s}, \\
      0\rule{0pt}{3ex} & \text{otherwise},
    \end{array} \right.
\end{equation}
where $\theta_\text{n}$ and $\theta_\text{s}$ refer to the northern and southern boundaries of the cellular flow, respectively.
The parameter $\alpha$ controls the skewness of the latitudinal profile, that is, the location of the peak velocity in latitude.

As for the radial profile of the flow model, we determine the $F(r)$ by assigning a few fixed points $\big(r_i, F(r_i)\big)$ and connecting them by cubic spline interpolation to have a smooth profile, where the index $i$ labels the fixed points.
The largest and smallest values of $r_i$ give the top and bottom boundaries of the cellular flow model (denoted by $r_\text{t}$ and $r_\text{b}$, respectively) beyond which the flow field is set to zero.
These fixed points are given in such a way that the $F(r)$ reverses its sign at least once along $r$.
The characteristic depth where the $F(r)$ changes sign, namely the reversal of the horizontal flow direction, is denoted by $r_\text{c}$.
The only constraint imposed on the radial profile is that the $f(r)$, obtained by integrating Eq.~\ref{eq:df} over $r$, must be zero at $r_\text{t}$ and $r_\text{b}$.
The strategy to determine the fixed points such that the $f(r)$ vanish at the boundaries is described in Appendix~\ref{app:knots}.

In this work the modeled meridional circulation is composed of two components, the global-scale meridional circulation (MC) that is present at all latitudes and the local cellular flows (LC) that is confined to the mean active latitudes.
The MC and LC models are constructed separately within the above-mentioned framework and then added up to form the final flow field.
The characteristics of flow models implemented here are listed in Table~\ref{tab:parm}.
Figure~\ref{fig:ur+uh} shows the latitudinal and radial profiles of these flow models.

We consider three MC models, MC1, MC2, and MC3, all of which have a common latitudinal profile but different radial profiles.
The radial profile of model MC1 resembles a simple structure in which the poleward meridional flow penetrates into the convection zone to a depth of 0.8~$R_\odot$ and reverses direction in the lower convection zone.
An early result of helioseismic inversion of MDI data with the constraint of mass conservation showed such a structure \citep{Giles2000}.
The radial profile of model MC2 still has a single-cell structure, but the flows in the upper convection zone decline more rapidly with increasing depth and maintain a rather weak poleward flow in the middle of the convection zone.
Besides, its characteristic depth $r_\text{c}$ is slightly deeper than 0.8~$R_\odot$.
Some numerical simulations based on mean-field hydrodynamics obtained this profile for the solar case \citep[e.g.,][]{Rempel2005,Kitchatinov2011}.
Also, a recent result of helioseismic inversion of HMI data exhibited a similar structure \citep{Rajaguru2015}.
While models MC1 and MC2 have a single-cell structure, model MC3 has a double-cell structure in which a shallow equatorward flow lies in the middle of the convection zone with poleward flows in the upper and lower convection zone \citep{Zhao2013,Chen2017,Lin2018}.
This type of radial profile was achieved by recent mean-field hydrodynamic models \citep{Bekki2017,Pipin2018}.

On the other hand, we implement three LC models, LC1, LC2, and LC3, that represent the longitudinal-averaged near-surface flows converging toward the activity belts with return (divergent) flows in deeper layers.
They are much broader and weaker than the inflows associated with individual active regions.
Each LC model consists of two local cellular flows symmetrically located in the northern and southern hemispheres (centered at $\pm17\degr$ latitudes with a size of 80$\degr$).
The only difference among the three LC models is the anchoring depth of their return flows.
The return flow of model LC1 peaks at a depth of $\sim$0.9~$R_\odot$ \citep[taken from, e.g.,][]{Chou2001,Beck2002,Chou2005} while that of model LC2 peaks at a depth much closer to the surface \citep[taken from, e.g.,][]{Haber2004,Zhao2004,Kosovichev2016}.
The third model, LC3, is anchored in the lower convection zone, which is not based on any observation but is implemented as a comparison to see how sensitive the travel-time shift is to a deeply-rooted local flow.

Because the eventual forward-modeled travel-time shift results from a sum of the MC and LC models, there is considerable flexibility in choosing the values of parameters provided that the net result matches the measurement.
In practice, we first match approximately the horizontal flow profile of the MC models in the near-surface layer to the surface observations \citep[e.g.,][, Fig.~11]{Hathaway2011a}, and then put in the LC models with some slight adjustments such that the forward-modeled travel-time shifts are compatible with the measured ones for the distance range $\Delta = 6\degr$--$12\degr$ (see the top row of Fig.~\ref{fig:dt-vs-lat}).\label{sec:A}
We remind the readers that these parameters are determined simply by eye rather than a fitting procedure, and not all the data points are used but the ones from shallower measurements because of better signal-to-noise ratio.

\subsection{Ray-approximation travel-time shifts}

\begin{figure}
  \resizebox{\hsize}{!}{\includegraphics{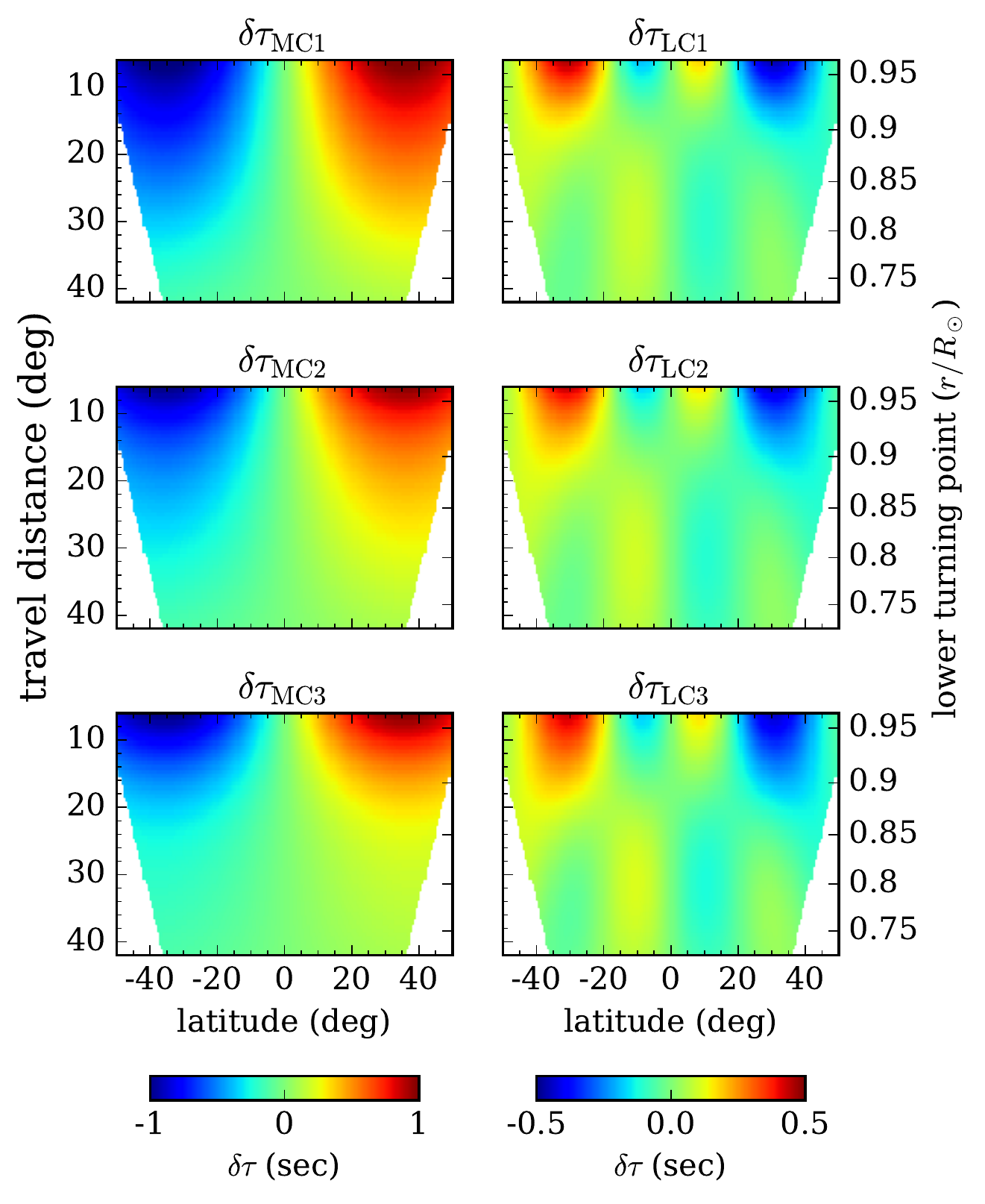}}
  \caption{ \label{fig:mc+lc,map}
    Ray-approximation travel-time shifts for the MC (left column) and LC (right column) flow models as a function of latitude and distance.
    The parts for which we do not make measurements are not shown in these maps.
    The color scale for each column is shown in the color bar at the bottom.
    The corresponding radii of the lower turning points are indicated on the right.
  }
\end{figure}

\begin{figure}
  \resizebox{\hsize}{!}{\includegraphics{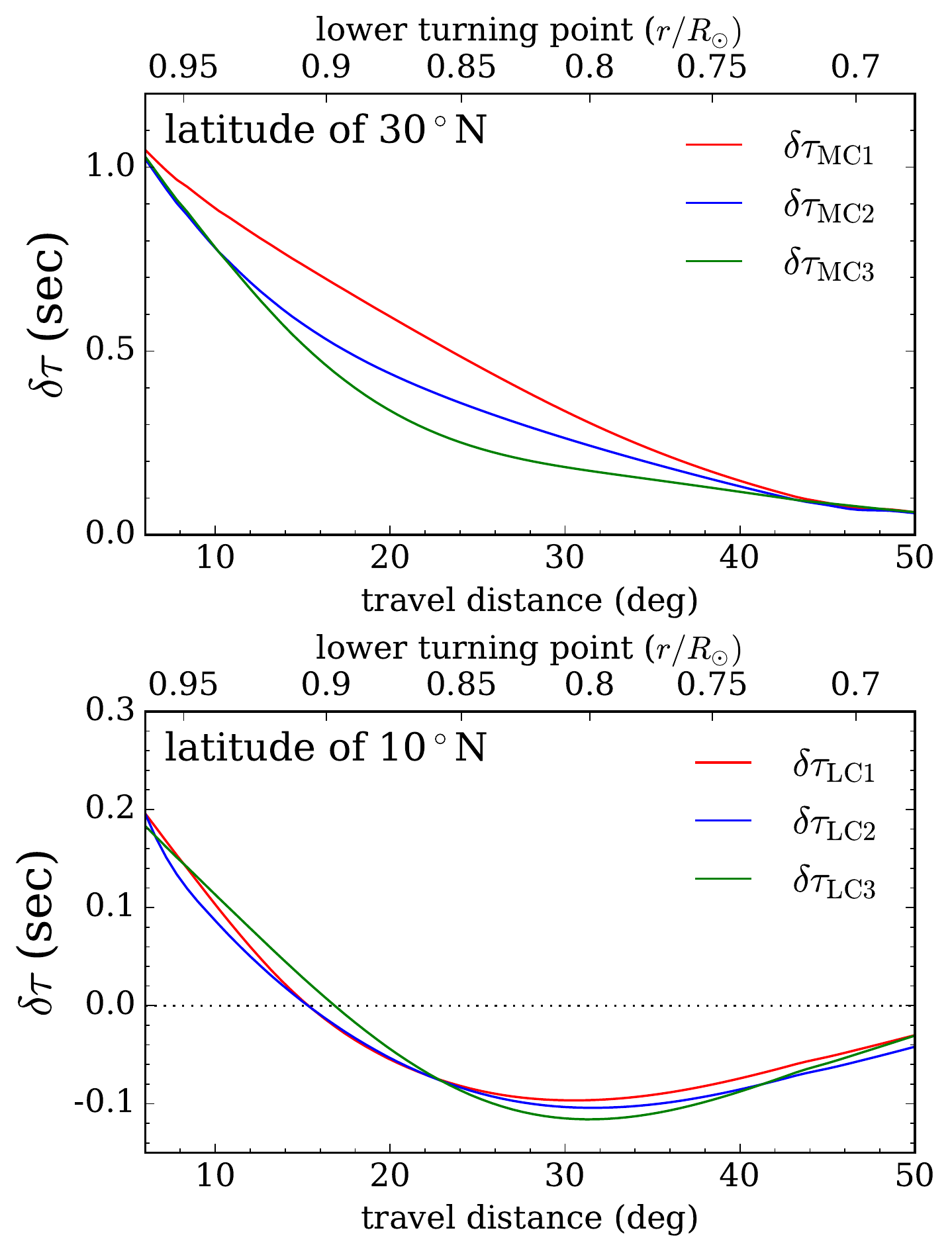}}
  \caption{ \label{fig:mc+lc,cf}
    \emph{Top}: cuts at the latitude of 30$\degr$ north through the $\delta\tau_\text{MC}$s as a function of distance.
    \emph{Bottom}: cuts at the latitude of 10$\degr$ north through the $\delta\tau_\text{LC}$s as a function of distance.
    The corresponding radii of the lower turning points are indicated at the top.
  }
\end{figure}

In the ray approximation \citep{Kosovichev1996,Kosovichev1997b}, the travel-time shift is estimated as an accumulation of travel-time perturbations caused by the flows along the ray path, which, incorporating an arc-to-arc geometry, gives
\begin{equation}\label{eq:ray}
  \delta\tau(\lambda, \Delta) = -\frac{2}{N} \sum_{i=1}^{N} \int_{\Gamma_i (\lambda, \Delta)} \frac{1}{c^2} \, \vec{u}\cdot\mathrm{d}\vec{l},
\end{equation}
where the line integral is calculated along the ray path $\Gamma_i$ connecting a pair of points that belong to the arc-to-arc geometry, $N$ is the number of ray paths, and $c$ is the sound speed from the solar model S \citep{Christensen-Dalsgaard1996}.
A frequency of 3.2~mHz, around which the filtered power spectrum peaks, is used for computing the ray paths.
The arc-to-arc geometry whereby the end points of multiple rays are arranged consists of two opposing arcs aligned in the north-south direction, separated by an angular distance $\Delta$, and centered at the latitude $\lambda$ (same geometry as for the measured $\delta\tau_\text{sn}$).
A comparison between single-ray and arc-averaged travel-time shifts as well as a schematic plot of the arc-to-arc geometry are shown in Appendix~\ref{app:arc}.

\label{sec:mc}
The ray-approximation travel-time shifts derived from the MC and LC models, denoted by $\delta\tau_\text{MC1}$, $\delta\tau_\text{LC1}$, etc., are shown in Fig.~\ref{fig:mc+lc,map}.
Cuts through the $\delta\tau_\text{MC}$s and $\delta\tau_\text{LC}$s, are also shown in Fig.~\ref{fig:mc+lc,cf}.
Clearly, the result of the $\delta\tau_\text{MC}$s shows that the extent and the decreasing rate of the poleward flows in the upper part of the convection zone determines how fast the forward-modeled travel-time shifts decline with increasing travel distance.
This result is qualitatively in agreement with that of \citet[, Figs.~3.5 and 3.10]{Chakraborty2015} in which the shallower the return flows in a flow model, the steeper the decline in the derived travel-time shift with increasing travel distance.
Besides, the flows in the lower convection zone seem to make little contribution to the travel-time shifts for $\Delta > 30\degr$.
By examining the line integral in Eq.~\ref{eq:ray} for contributions from the line segments in the lower convection zone, we find a deep flow there with a peak value of 3~m~s$^{-1}$ only produces a travel-time shift less than 0.05~s, which is consistent with the estimate made by \citet{Braun2008}.
That is to say, the forward-modeled travel-time shifts derived from rays whose lower turning points are in the lower convection zone contain a significant contribution from the flows in the upper part of the convection zone.
This explains why the $\delta\tau_\text{MC}$s do not change their signs as the distance increases regardless of the reverse of the flow direction.
The decline in the $\delta\tau_\text{MC}$s for $\Delta > 30\degr$ is probably owing to the fact that ray paths in the upper convection zone gradually become vertical with increasing travel distance.
In spite of the insensitivity to flows in the lower convection zone, the largest difference among the three $\delta\tau_\text{MC}$s is about 0.1--0.2~s in the regime $\Delta = 15\degr$--$30\degr$, which is, fortunately, discernible with a careful measurement.

That the travel-time shifts for large-distance cases are influenced by the flows in the upper convection zone becomes more apparent for local flows.
This can be seen in the bottom panel of Fig.~\ref{fig:mc+lc,cf} where the $\delta\tau_\text{LC1}$ and $\delta\tau_\text{LC2}$ persist for $\Delta > 30\degr$ despite the fact that there are no flows below 0.8~$R_\odot$ in models LC1 and LC2.
Also, all three $\delta\tau_\text{LC}$s change signs around $\Delta = 15\degr$--$20\degr$ notwithstanding the diverse $r_\text{c}$ in the corresponding LC models.
In fact, the largest difference among the $\delta\tau_\text{LC}$s lies in $\Delta < 6\degr$, which is outside the scope of our measurement.
It seems difficult to resolve the anchoring depth of a local cellular flow for lack of an adequate near-surface measurement.
We note that the different solar cycle variations of the $\delta\tau_\text{sn}'$ for $\Delta < 18\degr$ and $\Delta > 18\degr$ as shown in the top two panels of Fig.~\ref{fig:dt-vs-t} might be partly due to the sign change of the $\delta\tau_\text{LC}$ around $\Delta = 18\degr$.

\begin{figure*}[!h]
  \sidecaption
  \includegraphics[width=12cm]{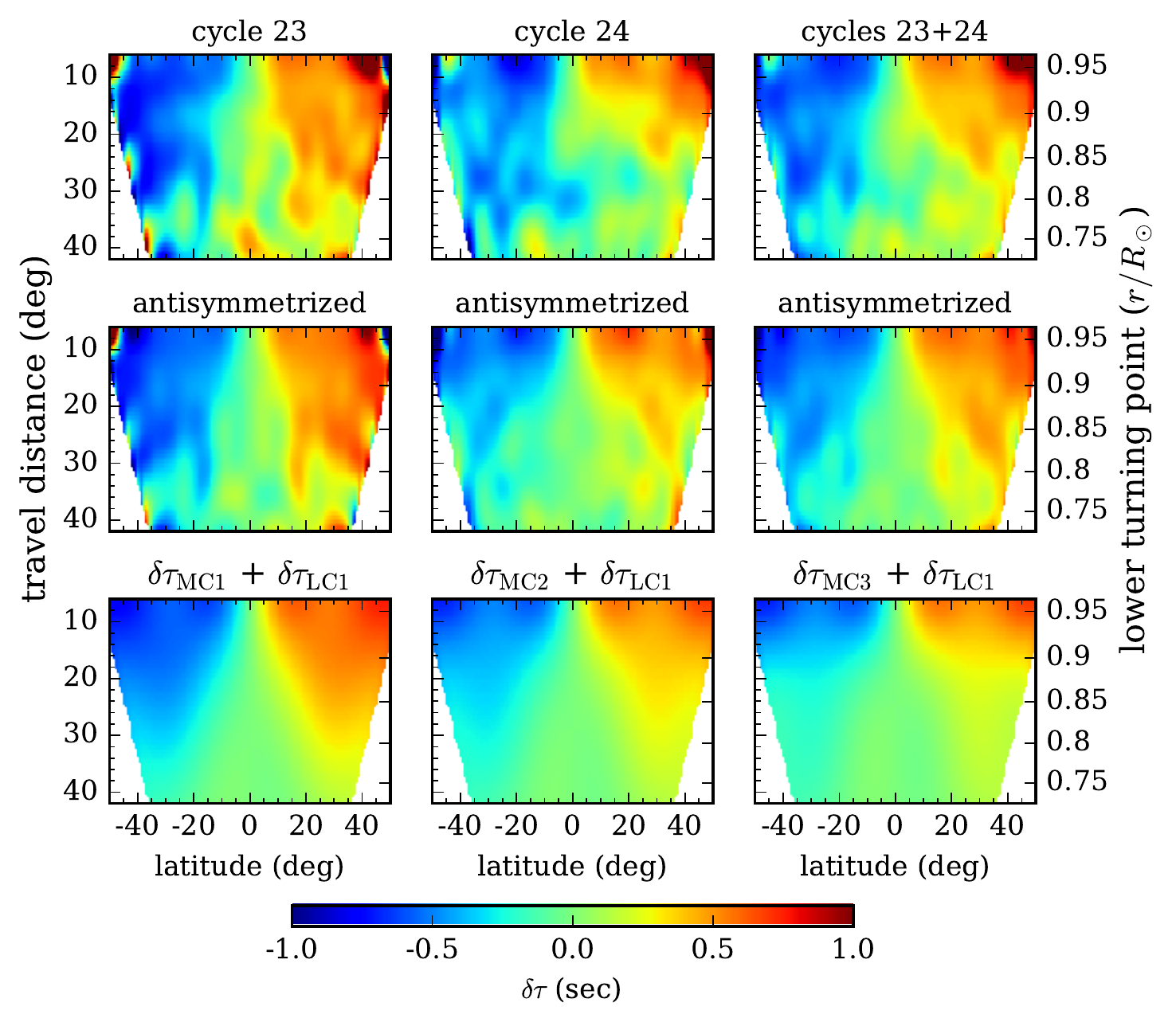}
  \caption{ \label{fig:dist-vs-lat}
    Measured and forward-modeled travel-time shifts as a function of distance and latitude.
    The top row shows maps of the three measured travel-time shifts, $\delta\tau_\text{sn;23}'$, $\delta\tau_\text{sn;24}'$, and $\delta\tau_\text{sn;23+24}'$, from left to right.
    The middle row shows the same $\delta\tau_\text{sn}'$s as the top row but antisymmetrized about the equator.
    The bottom row shows the forward-modeled travel-time shifts derived from the MC1, MC2, and MC3 flow models, from left to right, in combination with that from model LC1.
    All maps of the measured and forward-modeled travel-time shifts are Gaussian smoothed with a FWHM of 7.2$\degr$ in latitude and in distance.
    The data used to plot the top row are available at the CDS.
  }
\end{figure*}

Furthermore, we try an LC model (not shown here) with a smaller size of the cellular flows but otherwise the same configuration as the LC1 model.
The magnitude of the resulting $\delta\tau_\text{LC}$ at low latitudes is, instead, greater than that of the $\delta\tau_\text{LC1}$ owing to less overlapping and thus less cancellation of the cellular flows in the opposite hemispheres.
This implies that an asymmetry between the opposite cells at active latitudes may easily lead to a non-zero $\delta\tau_\text{LC}$ at the equator, which might be confused with a $P$-angle-error-induced time shift as discussed in Sect.~\ref{sec:dP}.
We also note that the magnitude of the $\delta\tau_\text{LC}$s are still on the order of 0.1~s for $\Delta > 30\degr$, which is comparable to that of the $\delta\tau_\text{MC}$s.
The pattern in $\delta\tau_\text{LC}$s beyond the distance of 30$\degr$ alternating in latitude (as shown in the right column of Fig.~\ref{fig:mc+lc,map}) might therefore be misinterpreted as a change in the deep meridional flows as the near-surface inflows migrate toward the equator with time.
Apparently, these near-surface inflows add a further complication to the interpretation of the deep meridional flow measurement in addition to the systematic errors.

\section{Comparison of observed and modeled travel times} \label{sec:cmpr}

Figure~\ref{fig:dist-vs-lat} gives an overview of the three long-term averaged travel-time shifts, $\delta\tau_\text{sn;23}'$, $\delta\tau_\text{sn;24}'$, and $\delta\tau_\text{sn;23+24}'$, with and without antisymmetrization, together with three forward-modeled results, $\delta\tau_\text{MC1}+\delta\tau_\text{LC1}$, $\delta\tau_\text{MC2}+\delta\tau_\text{LC1}$, and $\delta\tau_\text{MC3}+\delta\tau_\text{LC1}$.
Since the difference among the three $\delta\tau_\text{LC}$s are within the error bars, only the $\delta\tau_\text{LC1}$ are shown in this section for clarity.
In the top row of Fig.~\ref{fig:dist-vs-lat}, unlike the $\delta\tau_\text{sn;23}'$, the $\delta\tau_\text{sn;24}'$ shows a strong north-south asymmetry for distances greater than 12$\degr$, indicating the peculiarity of cycle 24.
The rapid decline in the northern $\delta\tau_\text{sn;24}'$ with increasing distance implies a rapid decrease of the poleward flows with increasing depth as illustrated in Sect.~\ref{sec:mc}.
In the middle row of Fig.~\ref{fig:dist-vs-lat}, the antisymmetric part of $\delta\tau_\text{sn;23}'$ seems to favor the result of $\delta\tau_\text{MC1}+\delta\tau_\text{LC1}$ whereas that of $\delta\tau_\text{sn;24}'$ favors the result of $\delta\tau_\text{MC2}+\delta\tau_\text{LC1}$.

To proceed, we shall make a detailed comparison of the travel-time shifts plotted as a function of latitude and as a function of travel distance.
Because of the complexities of asymmetric flow models, at the first attempt we only implement flow models that are antisymmetric about the equator; in that case, it is not meaningful to compare the asymmetric latitudinal profile of the measured $\delta\tau_\text{sn}'$ with the forward-modeled results.
Therefore, only the antisymmetric part of the latitudinal profiles is compared.
Otherwise, the $\delta\tau_\text{sn}'$ is averaged over a range of latitudes to have a qualitative comparison of the travel-time shifts as a function of travel distance.

\begin{figure*}
  \centering
  \includegraphics[width=17cm]{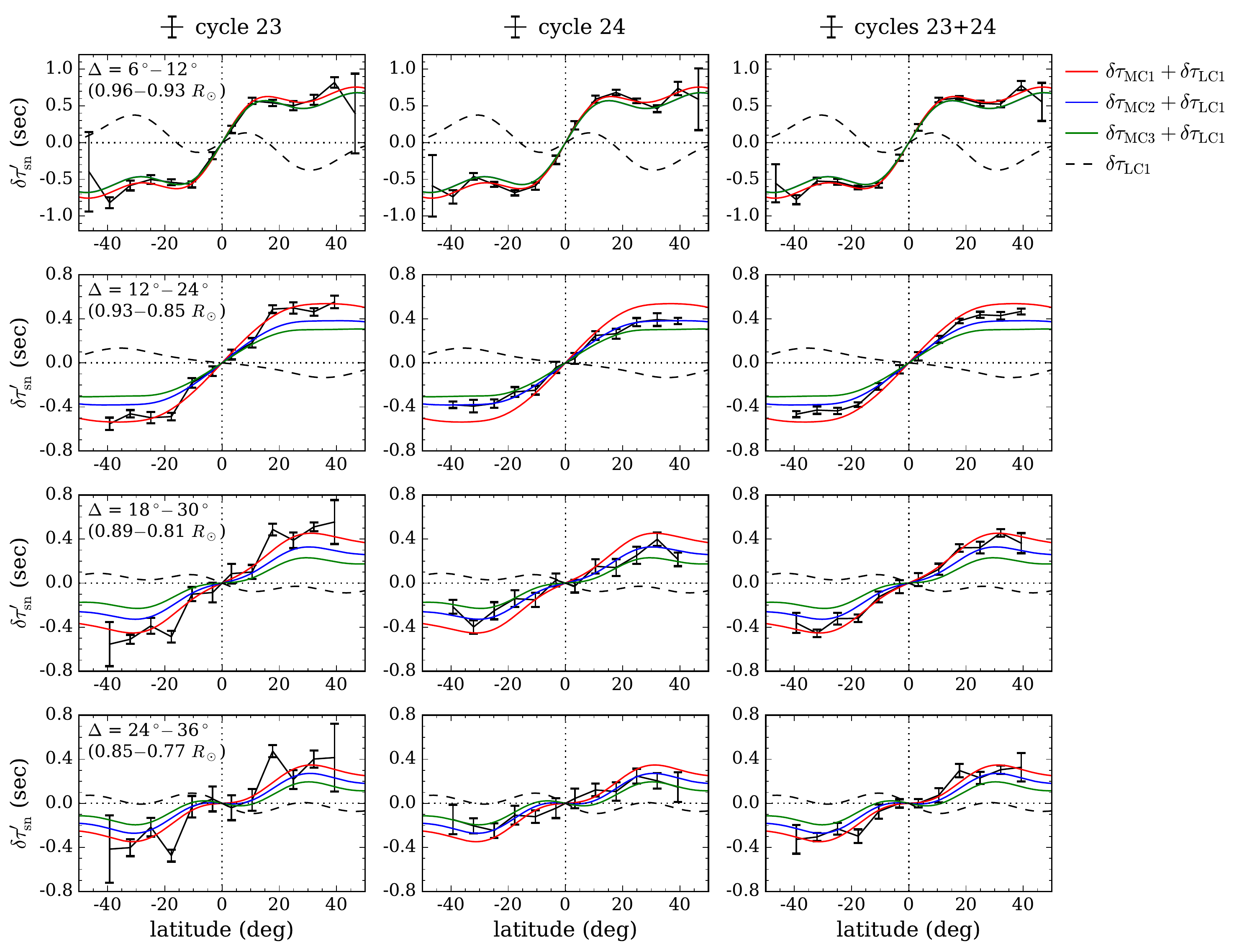}
  \caption{ \label{fig:dt-vs-lat}
    Three measured travel-time shifts (from left to right), $\delta\tau_\text{sn;23}'$, $\delta\tau_\text{sn;24}'$, and $\delta\tau_\text{sn;23+24}'$, averaged over four distance ranges and antisymmetrized about the equator, as a function of latitude (curves with error bars).
    Each row corresponds to a distance range indicated in the first column; shortest at the top and largest at the bottom.
    The data values are binned every 7.2$\degr$ in latitude.
    The error bars give the standard deviation of the mean in each binning interval.
    For comparison, we overplot the selected forward-modeled travel-time shifts (see the legend on the right) averaged over the same distance ranges as for $\delta\tau_\text{sn}'$s.
    As mentioned in Sect.~\ref{sec:A}, the parameters of the flow models were chosen such that the forward-modeled travel-time shifts are comparable to the measured $\delta\tau_\text{sn}'$s in the top row.
  }
\end{figure*}

\begin{figure}
  \resizebox{\hsize}{!}{\includegraphics{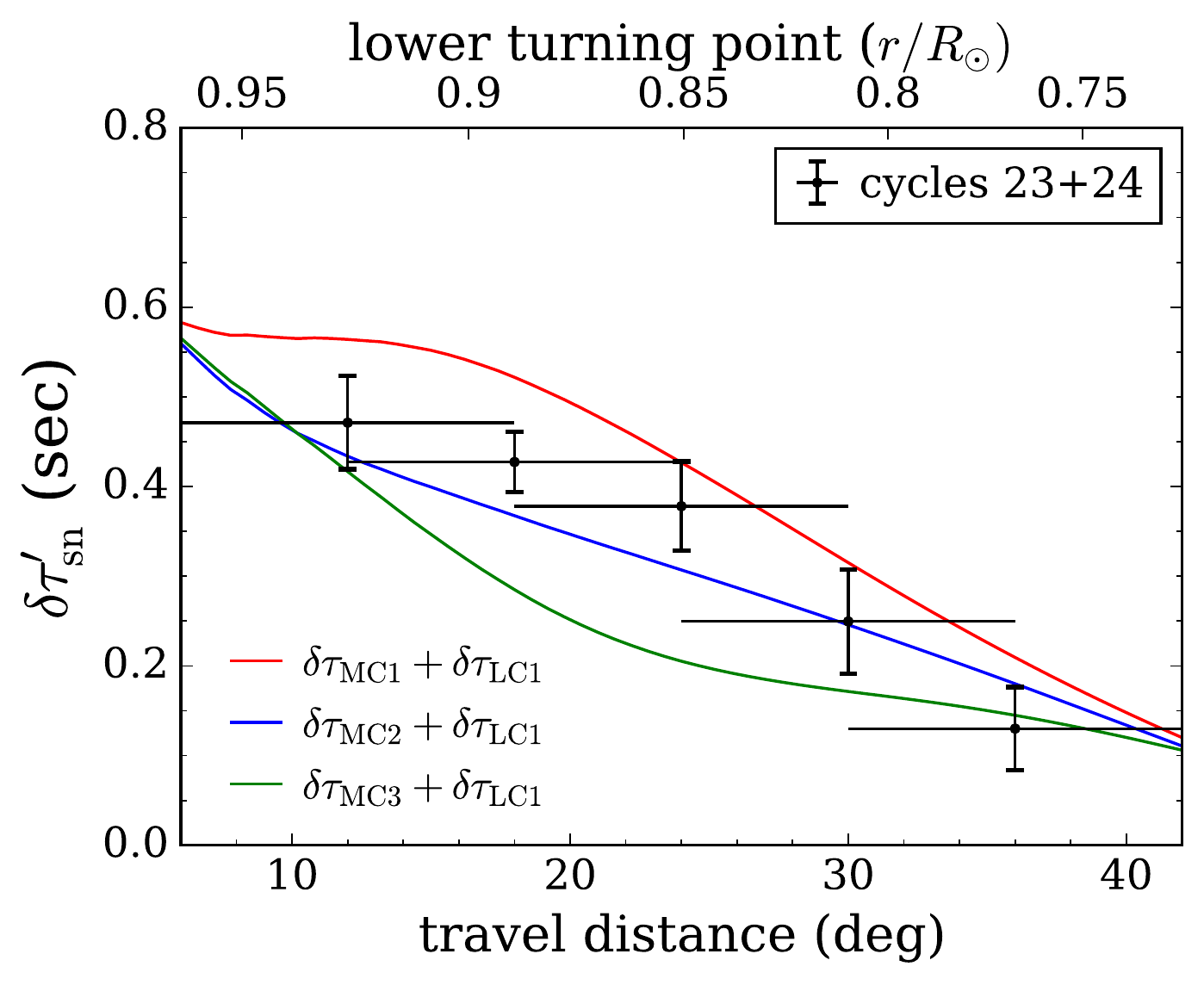}}
  \caption{ \label{fig:dt-vs-dist}
    Comparison of the 21-yr averaged travel-time shifts and the forward-modeled results as a function of distance.
    The data points with error bars are the $\delta\tau_\text{sn;23+24}'$ antisymmetrized about the equator and averaged over the latitude range 20$\degr$--35$\degr$ north as a function of distance.
    The horizontal bar associated with each point represents the distance range over which the mean value and the standard deviation of the mean are computed.
    The colored curves (see legend) indicate the selected forward-modeled travel-time shifts averaged over the same latitudinal range as for the $\delta\tau_\text{sn;23+24}'$.
    The models MC1 and MC2 are two varieties of single-cell flow models whereas the model MC3 has a double-cell structure (see Fig.~\ref{fig:ur+uh}).
  }
\end{figure}

\subsection{Antisymmetric part of $\delta\tau_\text{sn}'$}
Figure~\ref{fig:dt-vs-lat} shows a comparison between the antisymmetrized $\delta\tau_\text{sn}'$s and the forward-modeled results at four distance ranges.
In the top row, the forward-modeled results are compatible with the measurements within the errors as the model parameters were chosen to be so.
The slight mismatches between the measured and forward-modeled travel-time shifts for the latitudes of 10$\degr$--30$\degr$ as shown in the top-row left-two panels indicate the misalignment of the centroid position of the inflow models.
The characteristics of the longitudinal-average inflows depend on the equatorward drift rate of the active latitudes and hence are sensitive to the choice of the averaging period.
Because at the time of writing the minimum of cycle 24 has not yet reached and, in addition, half of the data from mid-2003 to mid-2010 are not used in the analysis, it would be inappropriate to make a direct comparison of the inflows with $\delta\tau_\text{sn;23}'$ and $\delta\tau_\text{sn;24}'$.
Despite that, we could still draw a qualitative comparison of the overall flow profiles between the two cycles.

Starting from the second row in Fig.~\ref{fig:dt-vs-lat}, the three forward-modeled results diverge from each other.
In the second row, it is obvious that the antisymmetrized $\delta\tau_\text{sn;23}'$ matches the $\delta\tau_\text{MC1}+\delta\tau_\text{LC1}$ while the $\delta\tau_\text{sn;24}'$ matches the $\delta\tau_\text{MC2}+\delta\tau_\text{LC1}$.
And the $\delta\tau_\text{sn;23+24}'$ just shows an averaged result between the $\delta\tau_\text{sn;23}'$ and $\delta\tau_\text{sn;24}'$.
The third row shows similar results as in the second row except that both the $\delta\tau_\text{MC2}+\delta\tau_\text{LC1}$ and $\delta\tau_\text{MC3}+\delta\tau_\text{LC1}$ are now in agreement with $\delta\tau_\text{sn;24}'$ within the errors.
In the bottom row, however, the differences among the forward-modeled results become comparable to the error bars.
In other words, we are unable to identify the meridional flow profile in the lower convection zone with the noise level in this work.

Although the antisymmetrized $\delta\tau_\text{sn;24}'$ favors an in-between model that could have a weak poleward flow or an equatorward flow in the middle of the convection zone, one should be careful not to jump to conclusions since the $\delta\tau_\text{sn;24}'$ is highly asymmetric for large distances.
The purpose of the comparison of the latitudinal profiles is to demonstrate the essential role played by the near-surface inflows in the interpretation of the travel-time measurement of the meridional flows, without which the $\delta\tau_\text{MC}$s would not match the overall profiles of the measured $\delta\tau_\text{sn}'$s in any way.

Apart from the latitudinal profile, another way to look at the data is to plot the travel-time shifts as a function of travel distance as shown in Fig.~\ref{fig:dt-vs-dist}.
The data points of the antisymmetrized $\delta\tau_\text{sn;23+24}'$ mostly fall in between the $\delta\tau_\text{MC1}+\delta\tau_\text{LC1}$ and $\delta\tau_\text{MC2}+\delta\tau_\text{LC1}$.
Since models MC1 and MC2 are just two varieties of single-cell flow models with different amplitudes of poleward flows in the middle of the convection zone, Fig.~\ref{fig:dt-vs-dist} suggests that the 21-yr averaged result generally favors a single-cell meridional circulation in the convection zone.

\begin{figure}
  \resizebox{\hsize}{!}{\includegraphics{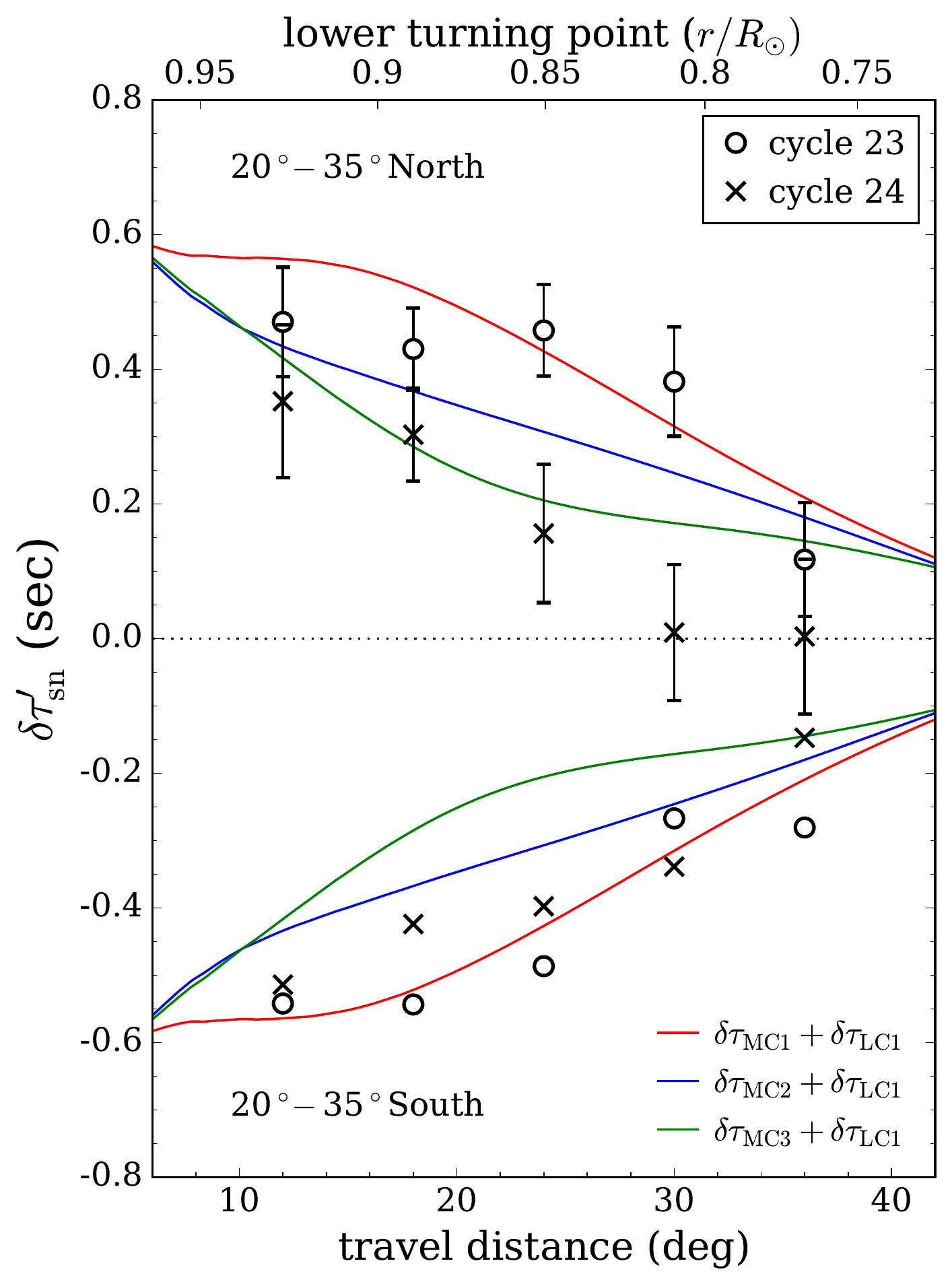}}
  \caption{ \label{fig:dt-vs-dist,hem}
    Same as Fig.~\ref{fig:dt-vs-dist} but for the $\delta\tau_\text{sn;23}'$ (open circles) and $\delta\tau_\text{sn;24}'$ (crosses) averaged over the latitude range 20$\degr$--35$\degr$ north and south separately.
    The horizontal bars in Fig.~\ref{fig:dt-vs-dist} are omitted here for clarity.
    The error bars of the southern $\delta\tau_\text{sn;23}'$ and $\delta\tau_\text{sn;24}'$ are similar to that of the northern ones and are omitted too.
    The models MC1 and MC2 are two varieties of single-cell flow models whereas the model MC3 has a double-cell structure (see Fig.~\ref{fig:ur+uh}).
  }
\end{figure}

\subsection{Comparison between the northern and southern $\delta\tau_\text{sn}'$}
Figure~\ref{fig:dt-vs-dist,hem} shows the $\delta\tau_\text{sn;23}'$ and $\delta\tau_\text{sn;24}'$ in the northern and southern hemispheres separately as a function of travel distance.
The most striking feature is the steep decline in the northern $\delta\tau_\text{sn;24}'$ with increasing travel distance.
While the northern $\delta\tau_\text{sn;23}'$ and $\delta\tau_\text{sn;24}'$ are still comparable for $\Delta < 12\degr$ (see the top panel of Fig.~\ref{fig:obs}), the two northern measurements substantially deviate from each other for $\Delta > 12\degr$, which is not the case at all for the southern measurements.
The southern measurements for the two cycles both agree with the single-cell flow models and do not show apparent cycle-to-cycle variation.
We note that the northern $\delta\tau_\text{sn;23}'$ is slightly smaller than the southern one for short distances; that is, the north-south asymmetry in the $\delta\tau_\text{sn;23}'$ is modest and is likely to be caused by near-surface phenomena.
In contrast, the north-south asymmetry in the $\delta\tau_\text{sn;24}'$ is surprisingly large, especially for large distances, and the decline in the northern $\delta\tau_\text{sn;24}'$ for large distances is faster than all the forward-modeled results.
With limited flow models implemented here, we cannot say too much about the behavior of the northern $\delta\tau_\text{sn;24}'$.

\section{Summary and discussion} \label{sec:sum}
We measured the travel-time shifts in the north-south and east-west directions from 14-yr of MDI data and 7-yr of HMI data covering 12-yr of cycle 23 and 9-yr of cycle 24.
The measured travel-time shifts exhibit several interesting features.
First, the temporal trends of the center-to-limb effect are different for the MDI and HMI data sets.
Second, the solar cycle variations of the travel-time shifts induced by the subsurface meridional flows are not in phase for different distance ranges.
Third, a significant reduction in the amplitude of travel-time shifts for large distances is seen in the northern hemisphere during the rising and maximum phases of cycle 24, which leads to an apparent cycle-to-cycle variation in the northern hemisphere and a strong north-south asymmetry for cycle 24.

The forward-modeled travel-time shifts were computed in the ray approximation for some representative flow models and compared with the long-term averages of measured travel-time shifts.
The 21-yr travel-time measurement in general favors the forward-modeled results for single-cell meridional circulation models in combination with the inflows toward the activity belts in the upper convection zone.
However, in view of the contrasts in the two cycles and in the two hemispheres, the northern measurement of cycle 24 decreases rapidly for distances greater than 12$\degr$, implying a rapid decrease of the poleward flows with increasing depth.
Due to limited flow models implemented in this work, we restrict ourselves from commenting further on the possible flow field that could produce such a rapid decline in the travel-time shifts.
Moreover, the forward-modeled results for the local inflows might partly explain the solar cycle variations of travel-time shifts for different distance ranges.

\paragraph{Systematic errors}
One might worry that the unusual result of cycle 24 in the northern hemisphere is caused by some systematic errors, especially when the center-to-limb variations observed by MDI and HMI are so different in many aspects.
We note that, after the removal of the center-to-limb effect, the results of \citet{Liang2017} measured from MDI and HMI observations in the period of one-year overlap (i.e., the rising phase of cycle 24) both showed a noticeable reduction in the northern hemisphere's measurements for $\Delta > 20\degr$.
However, we cannot exclude the possibility that this unusual reduction is caused by unknown systematic error if the systematic error affects both data sets in a similar way (a good example is the systematic effect caused by the surface magnetic field).
We also note that the strong north-south asymmetry of cycle 24, particularly the non-zero values at the equator, might be attributed to a $P$-angle error and thus be removed.
Considering the problems posed by various systematic errors, it would be necessary in the future to carry out a comparison with helioseismic observations from the GONG++ data since 2001.

\paragraph{Deep meridional flows}
As for the flow profile below 0.8~$R_\odot$, we are hindered by the low signal-to-noise ratio from giving a conclusive result of the meridional circulation in the lower convection zone.
While a possible flow in the lower convection zone may produce a travel-time shift of a few hundredths of a second, the noise level in this work, after averaging over 21-yr data and a large range of latitudes and travel distances, is on the order of 0.1~s in the regime of concern.
Accordingly, a direct detection of deep meridional circulation with confidence would require an observation on the order of hundred years.
This estimate is consistent with that made by \citet{Braun2008} and by \citet{Hanasoge2009}.
A common way to boost the signal-to-noise ratio with limited observation in time-distance helioseismology is to apply a spatio-temporal filter in Fourier domain \citep[e.g.,][]{Kholikov2014}.
One concern is how to incorporate the filter with the masking procedure for this type of filters may blend the Doppler signals in the quiet Sun with that in the active regions.
Another approach is to average the travel-time shifts over a wider band in longitude at the expense of introducing systematic error.
Although the resulting measurement might contain the center-to-limb variation, the relative change in them can still be used to study the temporal variation of deep meridional circulation as done by \citet{Liang2015b}, assuming that the temporal change in the center-to-limb variation is smooth.

\paragraph{Local inflows}
In addition to the global-scale meridional circulation that covers all latitudes, the local cellular flows around the activity belts have a crucial role to play in explaining the meridional-flow-induced time shifts.
The presence of the near-surface inflows toward the active regions results in a time-shift pattern which alternates in latitude and in travel distance.
As a consequence, the equatorward migration of the near-surface inflows causes an apparent solar cycle variation in the travel-time measurements of meridional circulation.
These local flows in the upper convection zone, after averaging over decades of observations, may still produce a travel-time shift comparable to that by a deep meridional flow and complicates the interpretation of the measured travel-time shifts for large distances.
We note that \citet{Lin2018} adopted a simplified model to separate the inflows from the global-scale meridional circulation and obtained a slower meridional flow at solar maximum than that at minimum.
Apart from that, the anchoring depth of the inflow structure is of interest.
The divergent-flow-like travel-time perturbation peaking at a depth of $\sim$0.9~$R_\odot$ \citep{Chou2001,Beck2002,Chou2005} might be linked with the near-surface inflow to form a cellular flow structure.
However, it has been pointed out by \citet{Liang2015a} that these measurements were made without considering the influence of the surface magnetic field on the travel-time shifts (this issue is revisited in Appendix~\ref{app:nobmask}).
The travel-time shifts measured in this work, with the surface magnetic field being taken care of though, are insensitive to the anchoring depth of the inflow structure for the distance range implemented in our measurements.
To resolve a local cellular flow structure in the near-surface layers, higher spatial resolution Dopplergrams are required for the travel-time measurement in the distance range $\Delta < 6\degr$, which is absent from this work.

\paragraph{Other uncertainties}
Our data analysis did not take into account a potential error in the orientation of the solar rotation axis determined by the Carrington elements \citep{Beck2005,Hathaway2010}.
In the frame determined from the observations by \citet{Carrington1863}, this uncertainty may introduce an apparent flow leaked from the solar rotation on the order of $\sim$3~m~s$^{-1}$ northward in summer and southward in winter when the $B_0$ angle is small.
Thus, there would be a systematic effect if the northward and southward leaking flows did not cancel each other out in the periods over which the long-term averages were performed.
We note that the MDI data we discarded since mid-2003 all have a small $B_0$ angle and the systematic effect due to the error in the Carrington elements is expected to be minimal.
We also note that the ray approximation is expected to be inaccurate when applied to structures with length scale comparable to the first Fresnel zone \citep{Birch2001}.
Furthermore, we compute the ray paths with merely one frequency even though the central frequency of the wavelet has a dependence on travel distance.
On closer inspection, this may incur an uncertainty of 0.01--0.03~s, for short distances in particular (see Appendix~\ref{app:cf-freq}).
In this regard, a sophisticated inversion with kernels that take into account finite-wavelength effects \citep[e.g.,][]{Boening2017,Gizon2017,Fournier2018,Mandal2018} is needed to correctly decipher the meaning of the measured travel-time shifts.
It would be preferable to study the north-south asymmetry and other fine structures of meridional circulation by inversion instead of forward modeling.

\begin{acknowledgements}
We thank the referee for helpful comments that improved the quality of the
manuscript.
Z.C.L. thanks Robert Cameron and Dean-Yi Chou for useful discussions.
The HMI data used are courtesy of NASA/SDO and the HMI science team.
SOHO is a project of international cooperation between ESA and NASA.
The sunspot numbers are from WDC-SILSO, Royal Observatory of Belgium, Brussels.
The data were processed at the German Data Center for SDO (GDC-SDO), funded by the German Aerospace Center (DLR).
Support is acknowledged from the SpaceInn and SOLARNET projects of the European Union.
L.G. acknowledges support from the NYU Abu Dhabi Center for Space Science under grant no. G1502.
We used the workflow management system Pegasus funded by The National Science Foundation under OCI SI2-SSI program grant \#1148515 and the OCI SDCI program grant \#0722019.
\end{acknowledgements}

\bibliographystyle{aa}
\bibliography{ref}

\begin{appendix}

\section{Constructing the radial profiles of flow models}
\label{app:knots}
\begin{figure}[h]
  \resizebox{\hsize}{!}{\includegraphics{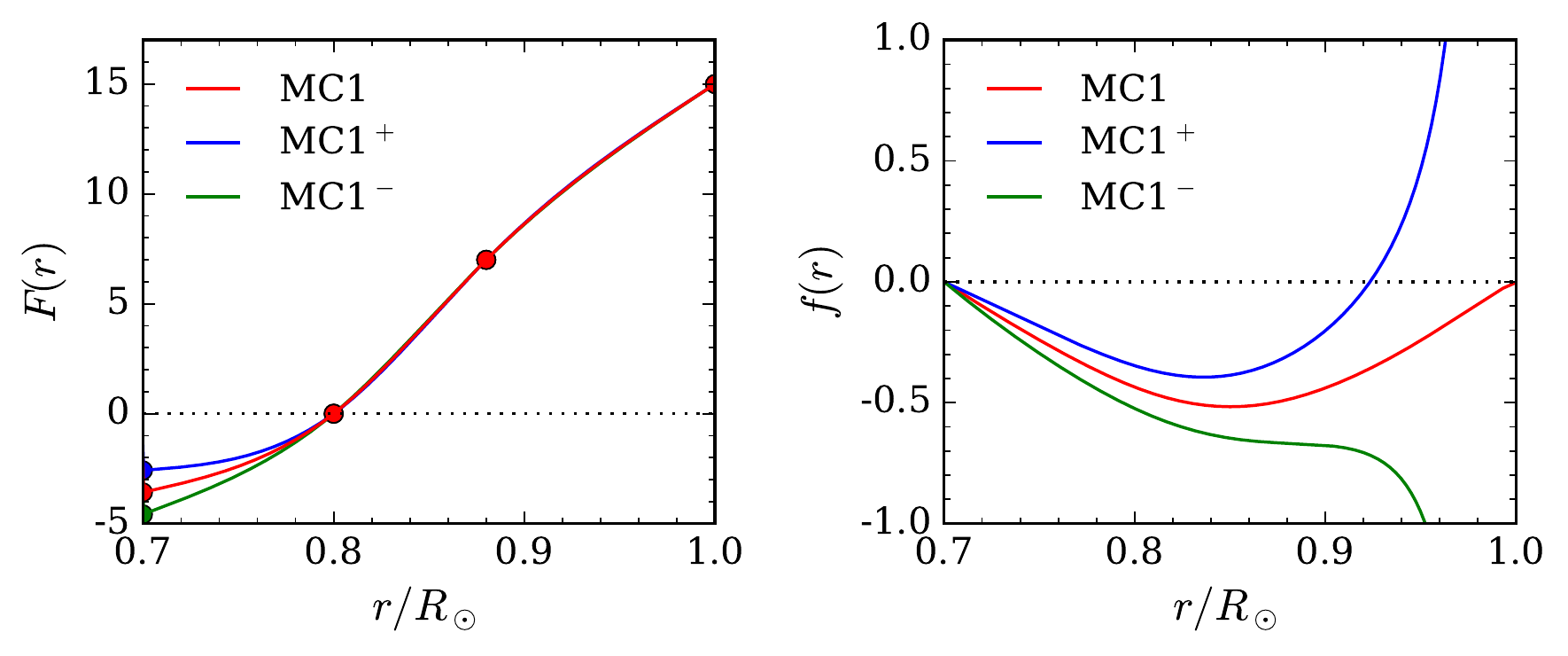}}
  \caption{ \label{fig:knots}
    Example of how an adjustment to the bottom point $\big(r_\text{b}, F(r_\text{b})\big)$ when constructing the $F(r)$ affects the resulting $f(r)$.
    The flow model used in the example is MC1 whose bottom is at $r_\text{b} = 0.7 R_\odot$.
    \emph{Left}: assigned points (filled circles) for constructing the $F(r)$ of model MC1 connected by cubic spline interpolation (red curve).
    The value of $F(r_\text{b})$ for model MC1 is deliberately increased (MC1$^+$, blue curve) or decreased (MC1$^-$, green curve) by one.
    \emph{Right}: resulting $f(r)$ from the three $F(r)$ in the left panel by computing Eq.~\ref{eq:f}.
    The values of $f(r_\text{t})$ for MC1$^+$ and MC1$^-$ are 25360.8 and -25360.8, respectively.
  }
\end{figure}

\begin{table}[h]
  \caption{Parameters for constructing the radial profiles of flow models} 
  \centering
  \begin{tabular}{ccl}
    \hline\hline
    Name\rule{0pt}{2ex} & $A_r$ & assigned points $\big(r_i/R_\odot, F(r_i)\,\big)$ \\
    \hline
    MC1 & 1.23 & (1, 15), (0.88, 7), (0.8, 0), (0.7, -3.58\tablefootmark{*}) \\
    \cline{3-3} 
    \multirow{2}{*}{MC2} & \multirow{2}{*}{1.23} & (1, 15), (0.98, 14), (0.95, 8), (0.87, 2), \\
                         & & (0.77, 0), (0.7, -3.11\tablefootmark{*}) \\
    \cline{3-3} 
    \multirow{2}{*}{MC3} & \multirow{2}{*}{1.23} & (1, 15), (0.98, 14), (0.915, 0), (0.8,  0), \\
                         & & (0.7,  0.15\tablefootmark{*}) \\
    \hline
    \multirow{2}{*}{LC1} & \multirow{2}{*}{-1.10} & (1, 10), (0.98, 9), (0.92, 0), (0.88, -0.7), \\
                         & & (0.81\tablefootmark{*}, 0) \\
    \cline{3-3} 
    \multirow{2}{*}{LC2} & \multirow{2}{*}{-1.85} & (1, 10), (0.99, 9), (0.97, 0), (0.96, -1), \\
                         & & (0.94\tablefootmark{*}, 0) \\
    \cline{3-3} 
    \multirow{2}{*}{LC3} & \multirow{2}{*}{-1.00} & (1, 10), (0.98, 9), (0.85, 0), (0.8, -1), \\
                         & & (0.72\tablefootmark{*}, 0) \\
    \hline
  \end{tabular}
  \tablefoot{ \label{tab:knots}
    \tablefoottext{*}{These numbers are determined by Newton method (see text) and are rounded in this table for clarity.
      We note that $f(r_\text{t})$ is very sensitive to these numbers as demonstrated in Fig.~\ref{fig:knots}; in that case, a precision of at least 6--8 decimal places for these numbers is needed to have $f(r_\text{t}) = 0$ within numerical errors.
    }
  }
\end{table}

First, we assign a few points $\big(r_i, F(r_i)\big)$ and connected them by cubic spline interpolation to form a desired profile of $F(r)$ for the horizontal flow component (defined in Eq.~\ref{eq:FG}).
These points are fixed except for the one at the bottom $\big(r_\text{b}, F(r_\text{b})\big)$, which will be adjusted later.
By integrating Eq.~\ref{eq:df} over $r$, we have the radial profile of the vertical flow component
\begin{equation} \label{eq:f}
  f(r) = \frac{1}{r^2\rho(r)}\int_{r_\text{b}}^{r} r'\rho(r') \, F(r') \, dr'.
\end{equation}

To confine the flows in a closed cell, the vertical flow is supposed to vanish at the boundaries; that is, $f(r_\text{b}) = 0$ and $f(r_\text{t}) = 0$.
We are free to set $f(r_\text{b})$ to zero.
As for $f(r_\text{t})$, we vary the location of the point at the bottom (i.e., the value of either $F(r_\text{b})$ or $r_\text{b}$) so that $f(r_\text{t}) = 0$.
This can be achieved by a one-dimensional root-finding method such as Newton's method.
Because the bottom of all MC models is fixed at the base of the convection zone (i.e., $r_\text{b} \equiv 0.7$~$R_\odot$), the $F(r_\text{b})$ is then adjusted to ensure $f(r_\text{t}) = 0$.
On the other hand, the horizontal flow at the bottom of the LC models is set to zero (i.e., $F(r_\text{b}) \equiv 0$) so the $r_\text{b}$ is adjusted instead of $F(r_\text{b})$.
Figure~\ref{fig:knots} shows an example of how an adjustment to the $F(r_\text{b})$ may affect the resulting $f(r_\text{t})$.
Table~\ref{tab:knots} gives the list of the assigned points as well as the $A_r$ for determining the radial profiles of all flow models in this work.

\newpage

\section{Comparison of arc-averaged travel-time shifts}
\label{app:arc}

\begin{figure}[h!]
  \resizebox{\hsize}{!}{\includegraphics{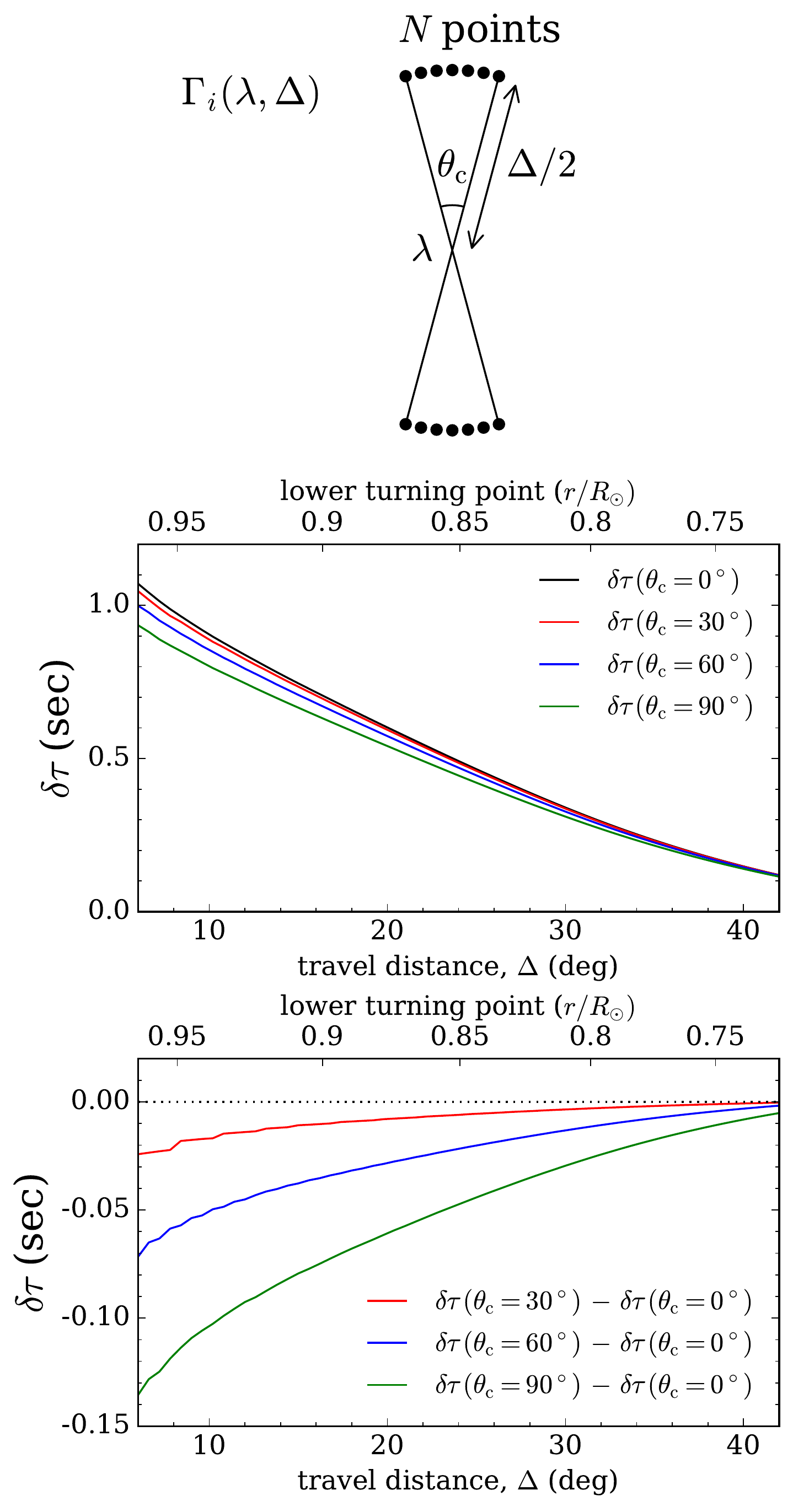}}
  \caption{ \label{fig:arc}
    Comparison of ray-approximation travel-time shifts derived from different arc-size configurations.
    \emph{Top}: schematic plot of the arc-to-arc geometry whereby the end points of multiple ray paths, $\Gamma_i(\lambda, \Delta)$, are arranged.
    The arc-to-arc geometry consists of two opposing arcs aligned in the north-south direction, separated by an angular distance $\Delta$, and centered at the latitude $\lambda$.
    The central angle subtended by an arc is denoted by $\theta_\text{c}$.
    The number of points on the arc, $N$, depends on the arc length.
    \emph{Middle}: single-ray and arc-averaged travel-time shifts cut at a latitude of 30$\degr$ north for different configurations in which the arc subtends central angles of 0$\degr$ (i.e., the single-ray case), 30$\degr$, 60$\degr$, and 90$\degr$, as a function of distance.
    The flow model used is MC1.
    \emph{Bottom}: differences between the single-ray travel-time shift and the other three arc-averaged travel-time shifts.
  }
\end{figure}

\newpage

\section{Revisiting the systematic effect introduced by the surface magnetic field}
\label{app:nobmask}

\begin{figure}[h!]
  \resizebox{\hsize}{!}{\includegraphics{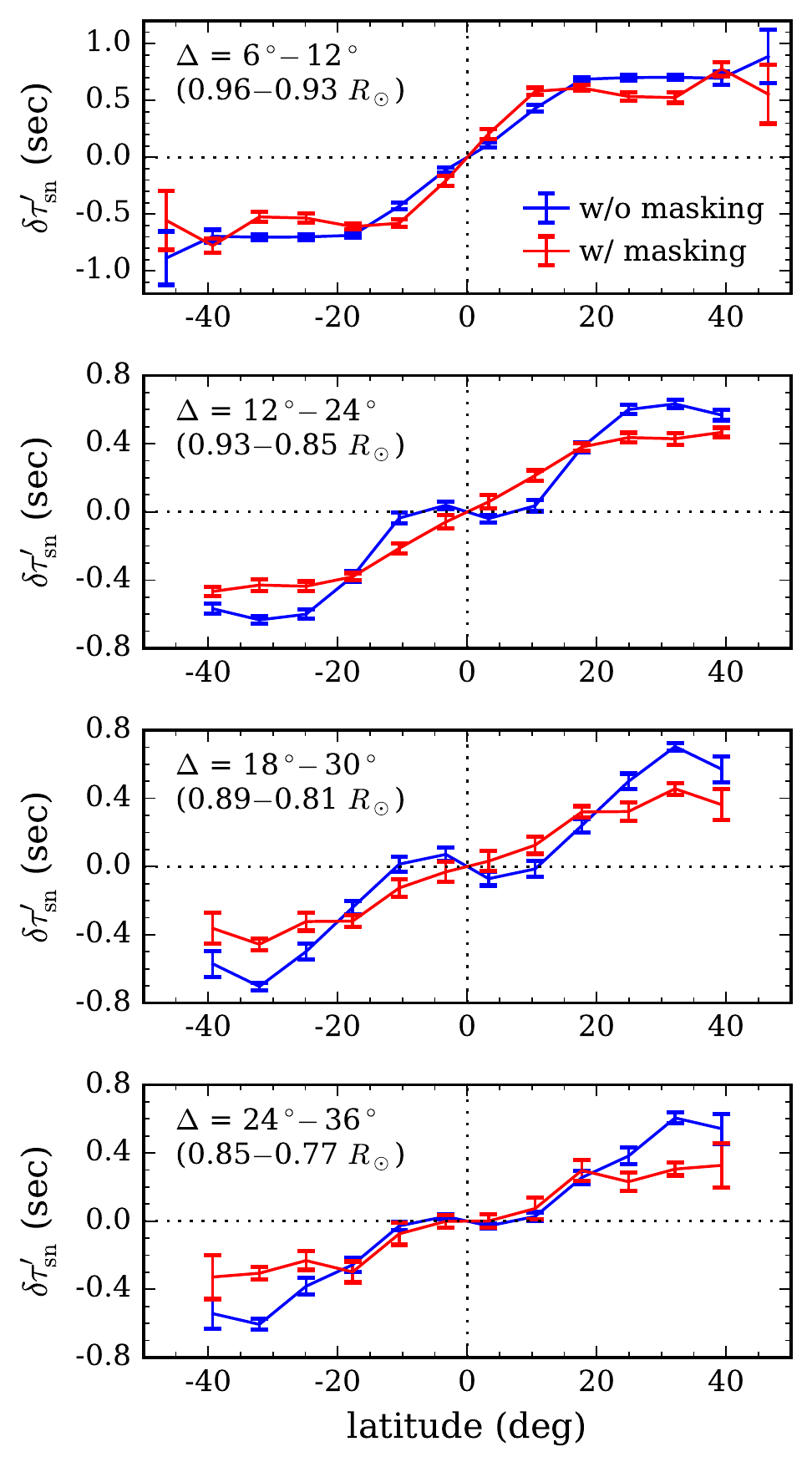}}
  \caption{ \label{fig:nobmask}
    Comparison of the measured $\delta\tau_\text{sn;23+24}'$ with (red) and without (blue) the masking procedure being used.
    The measured $\delta\tau_\text{sn;23+24}'$ with the masking procedure being used are the same as that shown in the right column of Fig.~\ref{fig:dt-vs-lat}.
  }
\end{figure}

At first, a time-varying pattern was found in the travel-time measurement of meridional circulation migrating toward the equator as the solar cycle evolves, suggesting a presence of divergent flows away from the active latitudes in a layer deep down to $\sim$0.9~$R_\odot$ \citep{Chou2001,Beck2002,Chou2005}.
Later, these presumed divergent flows were thought to be the counterpart of the near-surface inflows \citep{Zhao2004}.
However, \citet{Liang2015a} reported that this divergent-flow-like pattern in the travel-time measurement is mainly caused by the surface magnetic field.

To demonstrate how the surface magnetic field influences the travel-time measurements, we have repeated the data analysis for obtaining the $\delta\tau_\text{sn;23+24}'$ described in the main text except for the masking procedure.
The measured travel-time shift without the masking procedure being used is denoted by $\delta\tau_\text{sn;23+24}^{\prime\;\text{(w/o)}}$.
A comparison between the $\delta\tau_\text{sn;23+24}'$ and the $\delta\tau_\text{sn;23+24}^{\prime\;\text{(w/o)}}$ is shown in Fig.~\ref{fig:nobmask} \citep[cf. Fig.~7 in][, in which the center-to-limb effect had not been corrected]{Liang2015a}.
The systematic effect caused by the surface magnetic field can be clearly seen from this comparison, depending on the range of travel distances:
\begin{itemize}
  \item $\Delta = 6\degr$--$12\degr$: The latitudinal variation in the $\delta\tau_\text{sn;23+24}'$ due to the near-surface inflows is canceled out by this systematic effect and not seen in the $\delta\tau_\text{sn;23+24}^{\prime\;\text{(w/o)}}$ as if the inflows were to disappear at a depth shallower than 0.96~$R_\odot$.
  \item $\Delta = 12\degr$--$30\degr$: Since the inflow-induced variation becomes smooth in the $\delta\tau_\text{sn;23+24}'$, the magnetic-field-induced variation stands out in the $\delta\tau_\text{sn;23+24}^{\prime\;\text{(w/o)}}$ and resembles a divergent-flow-like travel-time perturbation.
  \item $\Delta > 30\degr$: Because the affected paired points used in the CCF computation are attached to the activity belts, the magnetic-field-induced variation in the $\delta\tau_\text{sn;23+24}^{\prime\;\text{(w/o)}}$ moves away from the mean active latitudes with increasing travel distance.
    Eventually, this variation cancels that from the opposite hemisphere at low latitudes as the central point of the affected paired points crosses the equator, but remains at higher latitudes.
\end{itemize}
To conclude, this magnetic-field-induced travel-time shift has an opposite sign to the inflow-induced travel-time shift for short-distance cases but a rather constant magnitude for all travel distances, implying that this systematic effect is probably caused by a phenomenon right beneath the sunspot or in the atmosphere above.
Otherwise, as pointed out by \citet{Gizon2008}, the presumed divergent flows inferred from the $\delta\tau_\text{sn;23+24}^{\prime\;\text{(w/o)}}$ for $\Delta = 12\degr$--$30\degr$ would be too large to balance the near-surface inflows concerning mass conservation.

We note that the magnitude of the travel-time shifts induced by the surface magnetic field is greater than that of the $\delta\tau_\text{LC}$s for $\Delta > 12\degr$.
This means, like the near-surface inflows, the surface magnetic field has an impact on the long-term averaged travel-time measurements of meridional circulation and must be taken into account in the analysis.

\newpage

\section{Frequency dependence of the forward-modeled travel-time shifts in the ray approximation}
\label{app:cf-freq}
\begin{figure}[h]
  \resizebox{\hsize}{!}{\includegraphics{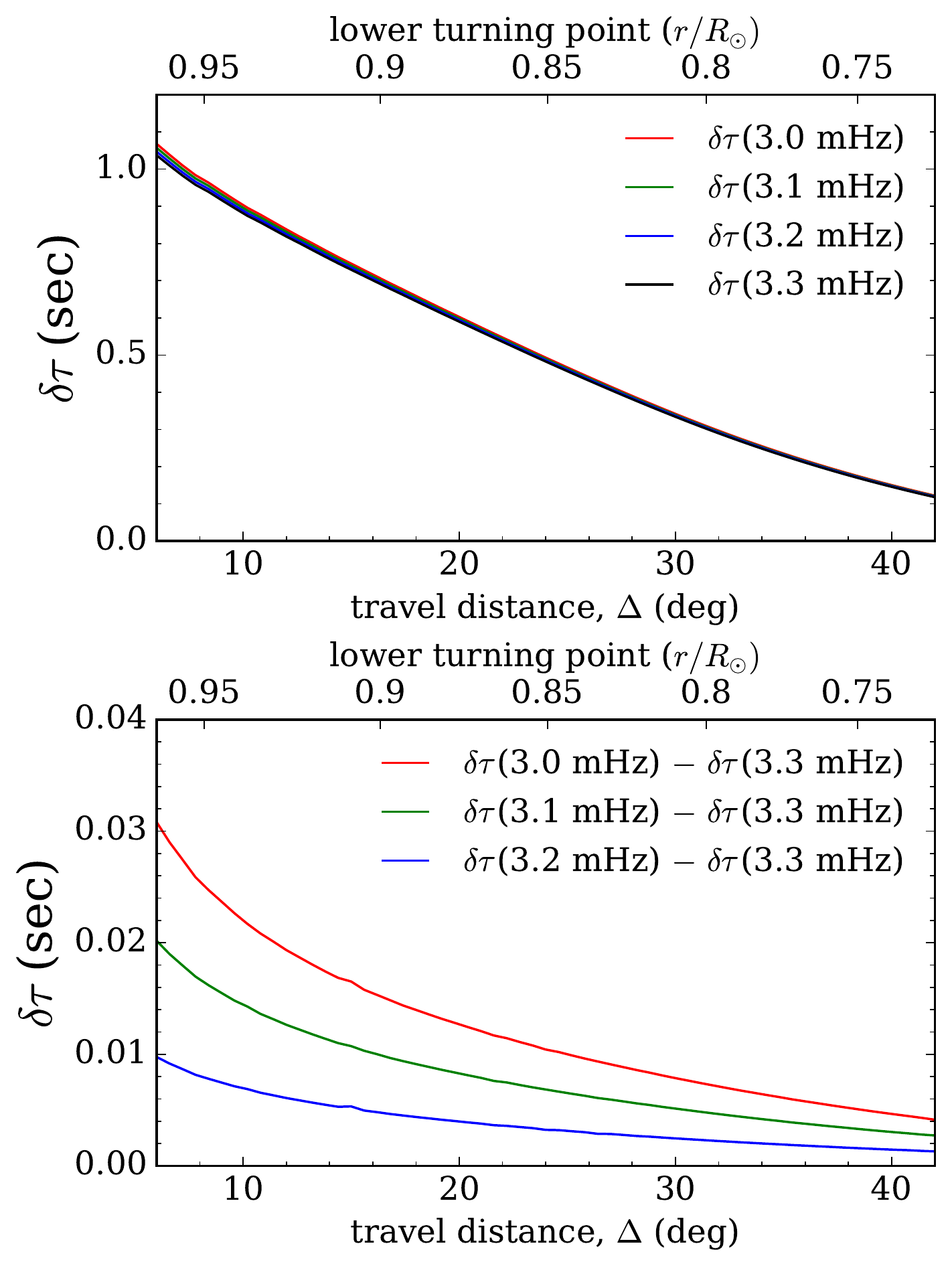}}
  \caption{
    \emph{Top}: comparison of ray-approximation travel-time shifts cut at a latitude of 30$\degr$ north computed from different frequencies as a function of distance.
    The frequencies used in this comparison are 3.0, 3.1, 3.2, and 3.3~mHz (see legend).
    A 30$\degr$-wide arc-to-arc geometry is implemented.
    The flow model used is MC1.
    \emph{Bottom}: differences between the travel-time shifts shown in the top panel.
    We note that the differences increase rapidly as the travel distance decreases, and are on the order of hundredths of a second for the shortest distance ($\Delta = 6\degr$) used in this work.
    Also, the smaller the frequency used in the ray approximation, the steeper the decline in the derived travel-time shift with increasing travel distance.
  }
\end{figure}

\end{appendix}

\end{document}